\documentclass[a4paper,fleqn,usenatbib]{mnras}

\usepackage{newtxtext,newtxmath}

\usepackage[T1]{fontenc}
\usepackage{ae,aecompl}

\usepackage{graphicx}	
\usepackage{amsmath}	
\usepackage{amssymb}	

\usepackage{subcaption} 
\defcitealias{Burles1998b}{BT} 

\title[D/H value at $z_{\rm abs} = 2.504$ towards Q1009$+$2956]{The primordial deuterium abundance at $z_{\rm abs}=2.504$ from a high signal-to-noise spectrum of Q1009$+$2956\\}

\author[E. O. Zavarygin et al.]{
E. O. Zavarygin,$^{1}$\thanks{E-mail: e.zavarygin@gmail.com (EOZ)}
J. K. Webb,$^{1,2}$
V. Dumont$^{3,1}$
and S. Riemer-S{\o}rensen$^{4,5}$
\\
$^{1}$School of Physics, University of New South Wales, Sydney, NSW 2052, Australia\\
$^{2}$DAMTP, Centre for Mathematical Sciences, Wilberforce Road, Cambridge, CB3 0WA, United Kingdom\\
$^{3}$Department of Physics, University of California, Berkeley, California 94720-7300, USA\\
$^{4}$Institute of Theoretical Astrophysics, The University of Oslo, Oslo NO-0316, Norway\\
$^{5}$ARC Centre of Excellence for All-sky Astrophysics (CAASTRO), NSW 2016, Australia
}

\date{Accepted XXX. Received YYY; in original form ZZZ}

\pubyear{2018}

\begin{document}
\label{firstpage}
\pagerange{\pageref{firstpage}--\pageref{lastpage}}
\maketitle

\begin{abstract}
The spectrum of the $z_{\rm em} = 2.63$ quasar Q1009$+$2956 has been observed extensively on the Keck telescope.  The Lyman limit absorption system $z_{\rm abs}=2.504$ was previously used to measure D/H by \citeauthor{Burles1998b} using a spectrum with signal to noise approximately 60 per pixel in the continuum near Ly\,$\alpha$ at $z_{\rm abs}=2.504$.  The larger dataset now available combines to form an exceptionally high signal to noise spectrum, around 147 per pixel.  Several heavy element absorption lines are detected in this LLS, providing strong constraints on the kinematic structure.  We explore a suite of absorption system models and find that the deuterium feature is likely to be contaminated by weak interloping Ly\,$\alpha$ absorption from a low column density H\,{\sc i} cloud, reducing the expected D/H precision.  We find D/H~$=2.48^{+0.41}_{-0.35} \times 10^{-5}$ for this system. Combining this new measurement with others from the literature and applying the method of Least Trimmed Squares to a statistical sample of 15 D/H measurements results in a ``reliable'' sample of 13 values. This sample yields a primordial deuterium abundance of $(\text{D/H})_{\rm p} = (2.545 \pm 0.025) \times 10^{-5}$. The corresponding mean baryonic density of the Universe is $\Omega_{\rm b}h^2 = 0.02174 \pm 0.00025$.  The quasar absorption data is of the same precision as, and marginally inconsistent with, the 2015 CMB Planck  (TT+lowP+lensing) measurement, $\Omega_{\rm b}h^2 = 0.02226 \pm 0.00023$.  Further quasar and more precise nuclear data are required to establish whether this is a random fluctuation.
\end{abstract}




\section{Introduction}

The primordial or Big Bang nucleosynthesis (BBN) epoch, a few minutes after the Big Bang, is the earliest time in the evolution of the Universe at which we are able to probe the physical properties of the universe at high precision \cite[see, e.g.,][and references therein]{Iocco2009,Cyburt2016}. It was an epoch where a handful of light elements, such as H, He, Li, Be, were synthesised. Of these elements, deuterium is known as the best baryometer due to its sensitivity to, and monotonic dependence on, the baryon-to-photon ratio $\eta_{10}$ or the baryon density $\Omega_{\rm b}$ \citep[since $\eta=273.9\Omega_{\rm b}h^2$,][]{Steigman2006}.

For years it has been known that the available D/H measurements show a large scatter relative to their uncertainties \citep[e.~g.][]{Balashev2016}. Whilst it is likely that this is simply a consequence of underestimating systematic errors, it is important to establish that is the case in order to rule out cosmological inhomogeneities. A few quasars that have previously been used for D/H estimates have subsequently been observed extensively such that the newer and much higher quality data permits a more accurate spectral analysis. The higher Signal-to-Noise (S/N) permits more stringent checks for  systematic problems in the data and allows more accurate modeling of the kinematic structure of the absorption system. 

The importance of high S/N data has recently been demonstrated by \cite{RiemerSorensen2015}. Their analysis of the $z_{\rm abs}=3.256$ absorption system towards PKS~1937$-$101 based on higher S/N data revealed a more complex kinematic structure than the previously published analysis \citep{Crighton2004}. The new measurement led to a higher value of D/H than the previous result and also led to a small reduction of the scatter. As another example, very recently \cite{RiemerSorensen2017} re-measured the $z_{\rm abs}=3.572$ absorption system towards PKS~1937$-$101 leading to a further slight reduction of the scatter. A further example is a systematic re-measurement of D/H in Damped Lyman Alpha systems by \cite{Cooke2014,Cooke2018}.
	
Here we present a re-measurement of the $z_{\rm abs}=2.504$ metal-poor Lyman limit system (LLS) towards the quasar Q1009$+$2956 (aka J1011$+$2941). Since the last analysis of this system by \cite{Burles1998b} (hereafter \citetalias{Burles1998b}) this quasar has been observed numerous times and the combined dataset yields an effective S/N of 147 in the continuum for Ly\,$\alpha$ at $z_{\rm abs}=2.504$, compared to about 60 for the \citetalias{Burles1998b} spectrum.

\section{Data reduction}
\label{sec:j1011_data_reduction}

\subsection{Data}

\begin{table*}
\begin{minipage}{\textwidth}
\caption{Journal for the Keck/HIRES observations of the Q1009$+$2956 used in this work.}
\label{tab:j1011_journal}
\begin{tabular}{@{}llllllll@{}}
	\hline
	Date       & PI  & Decker & Bin.$^a$ & Wavelengths$^b$, \AA & Res. power & $\sigma_v^c$, km s$^{-1}$ & Exposure, ks\\
	\hline
	1997-05-11 & Tytler  & C5 & 1 & $3160-4652$ & 37000 & 3.4 & 3         \\
	1998-12-14 & Tytler  & C5 & 1 & $3099-4619$ & 37000 & 3.4 & 8         \\                 
	1998-12-15 & Tytler  & C5 & 1 & $3099-4619$ & 37000 & 3.4 & 6.3       \\
	1999-03-08 & Tytler  & C5 & 1 & $3121-4642$ & 37000 & 3.4 & 7         \\
	1999-03-09 & Tytler  & C5 & 1 & $3121-4642$ & 37000 & 3.4 & 7.2       \\
	2003-06-07 & Sargent & C1 & 1 & $4605-7020$ & 49000 & 2.6 & $3+3$     \\
	2003-06-08 & Sargent & C1 & 1 & $4612-7033$ & 49000 & 2.6 & $3+3$     \\
	2004-11-05 & Tytler  & C5 & 1 & $3075-5883$ & 37000 & 3.4 & 3.6       \\
	2005-03-31 & Tytler  & C5 & 1 & $3129-5954$ & 37000 & 3.4 & 5.4       \\
	2005-04-30 & Sargent & C1 & 1 & $3157-5967$ & 49000 & 2.6 & $3+3$     \\
	2005-05-01 & Sargent & C1 & 1 & $3157-5967$ & 49000 & 2.6 & $3+3$     \\
	2005-05-31 & Steidel & C5 & 2 & $3138-5970$ & 37000 & 3.4 & $1.8+1.8$ \\
	2005-06-01 & Steidel & C5 & 2 & $3138-5970$ & 37000 & 3.4 & $1.8+1.8$ \\
	2005-12-06 & Tytler  & C5 & 1 & $3076-5897$ & 37000 & 3.4 & $6.4$	     \\
	2008-03-29 & Tytler  & C1 & 2 & $3129-5955$ & 49000 & 2.6 & $3.6+3.6$ \\
	\hline
	total & \multicolumn{6}{c}{} & 85.3 \\
	\hline
\end{tabular}

\small{$^a$CCD binning in the dispersion direction.}\\
\small{$^b$The wavelength values are in the vacuum/heliocentric frame.}\\
\small{$^c$Resolution in terms of $\sigma_v=\textsc{fwhm}/2\sqrt{2\ln2}$, where {\sc fwhm} is a full width at half maximum of the assumed Gaussian instrumental profile.}
\end{minipage}
\end{table*}

The quasar Q1009$+$2956 ($z_{\rm em}=2.63$) was observed with the Keck telescope using the High Resolution Echelle spectrograph \citep[HIRES,][]{Vogt1994} during many independent programs. The raw science and calibration data were provided by the W. M. Keck Observatory Archive (KOA)\footnote{\url{https://koa.ipac.caltech.edu/cgi-bin/KOA/nph-KOAlogin}}. Out of all the available exposures we ignored a small number with low S/N because cosmic ray identification and removal during data reduction is less effective in those cases, and because the impact on the final data quality was negligible. The dataset used in this study consists of 22 individual exposures with total observation time of 23.7 hours. Information about individual exposures is presented in Table~\ref{tab:j1011_journal}.

The HIRES data used by \citetalias{Burles1998b} (six exposures, see their table~1) are not included in our dataset. The reason for that is as follows. The first exposure (9000 sec, 28 Dec 1995) does not have a ThAr frame for wavelength calibration. Reduction of the second exposure (7200 sec, 28 Dec 1995) fails due to overlapping orders. The third exposure (4800 sec, 28 Dec 1995) has very low S/N and thus was not included in our dataset. Data of the remaining three exposures were not available from the archive or elsewhere.  Therefore, the estimate of the primordial deuterium abundance given in this paper is based on a completely independent dataset from \citetalias{Burles1998b}.

\subsection{Reduction}
\label{sec:j1011_reduction}
For data reduction, MAuna Kea Echelle Extraction (MAKEE) package\footnote{\url{http://www.astro.caltech.edu/~tb/makee/index.html}} developed by Tom Barlow has been used. A standard  data reduction procedure has been followed. This includes bias subtraction using a bias frame, tracing the object position on each CCD exposure using an image of either a bright star or a ``pinhole'' quartz lamp exposure, optimal extraction of the science spectra for each echelle order, flat-fielding using a quartz lamp frame, and wavelength calibration using a ThAr frame. For each science exposure the available calibration files closest in time and in setups were selected. MAKEE provides wavelength-calibrated spectra of each echelle order in the heliocentric coordinate frame. There is a known issue with the air-to-vacuum correction formula in MAKEE described in section~2.9 of \cite{Murphy2001}. We do therefore not apply this correction in MAKEE. In Appendix~\ref{app:reduction_issues} we describe all the specific data reduction issues and applied solutions for the data in Table~\ref{tab:j1011_journal}.

\begin{figure*}
\includegraphics[width=\linewidth]{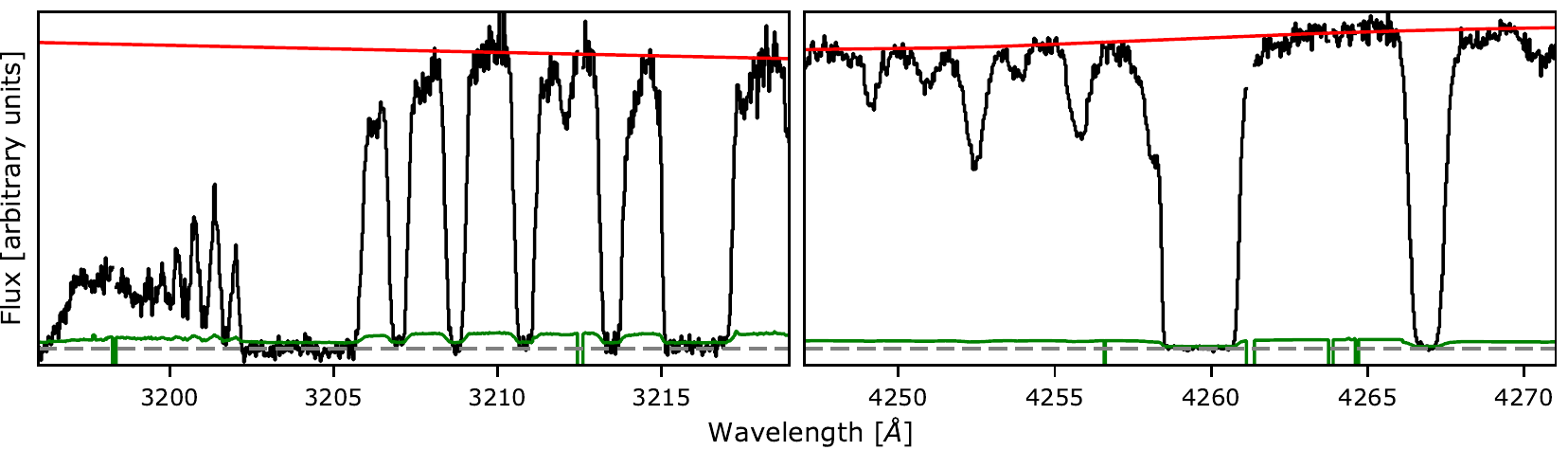}
\caption{A reduced non-flux-calibrated spectrum and the initial continuum estimate for Q1009$+$2956. The spectrum shown is the exposure obtained on 29 March 2008 (starting time 06:40, UT). The other exposures are similar.
The left and right panels show the Lyman limit and the Ly\,$\alpha$ regions of the $z_{\rm abs}=2.504$ LLS, respectively. The black histogram shows the flux in arbitrary units. The smooth red line on top of the flux indicates the initial continuum estimate. The horizontal dashed line shows a zero level.
The green histogram near zero shows $1\sigma$ error with negative values being pixels clipped due to cosmic rays.
}
\label{fig:2_504_cont_norm}
\end{figure*}

Subsequent processing of the spectra, including the air-to-vacuum correction, has been done in UVES\textunderscore popler\footnote{\url{http://astronomy.swin.edu.au/~mmurphy/UVES_popler/}} software \citep{MurphyUVES2016}. This software re-disperses spectra of echelle orders onto a common vacuum heliocentric wavelength scale, combines the echelle orders, and removes leftover cosmic rays by $\sigma$-clipping. The automatic procedure of cosmic ray clipping was checked by visual inspection. We also excluded all identified affected parts of the spectra (by blemishes of the CCD such as e.g. bad columns, bleeding regions etc., see Appendix~\ref{app:reduction_issues}). Initial continuum estimation was done in UVES\textunderscore popler. First, the spectra were divided into overlapping chunks of $10000-20000$~km~s$^{-1}$ in the Ly\,$\alpha$ forest and $2000-2500$~km~s$^{-1}$ in the red part, depending on S/N and setups. Then, the continuum in each chunk was fitted with low-order Chebyshev polynomial (fourth order at forest and sixth-eighth order in the red part, depending on S/N and setups). This automatic procedure was then checked by visual inspection to re-fit the continuum in places where the automatic procedure failed (mostly in the forest). Continuum across the high-order Lyman series transitions (shortwards of 3209~{\AA}) where flux does not return to its initial position is estimated as a 1-2-order (depending on S/N and setups) polynomial fit to the absorption-free regions redwards of the Lyman Limit, $3209-3249$~{\AA}, extrapolated to lower wavelengths. Fig.~\ref{fig:2_504_cont_norm} illustrates the continuum normalisation results. The whole process results in a sample of 22 one-dimensional normalised spectra, with a few spectral gaps due to either gaps between CCD chips or bad regions having been masked out.

\subsection{Velocity offsets and co-added spectra}
\label{sec:offset}

Previous analyses show that high-resolution echelle spectra may show velocity offsets between different exposures \citep[e.\,g.][]{Whitmore2010, Balashev2016} and long-range wavelength distortions within each exposure \citep[e.\,g.][]{Molaro2008, Rahmani2013, Whitmore2015, Dumont2017}. Whilst long-range distortion effects are important for varying constant analyses, they are unlikely to impact on the D/H measurement studied here so are ignored.  We do however correct for velocity shifts between individual exposures.

\begin{figure}
\center
\includegraphics[width=\linewidth]{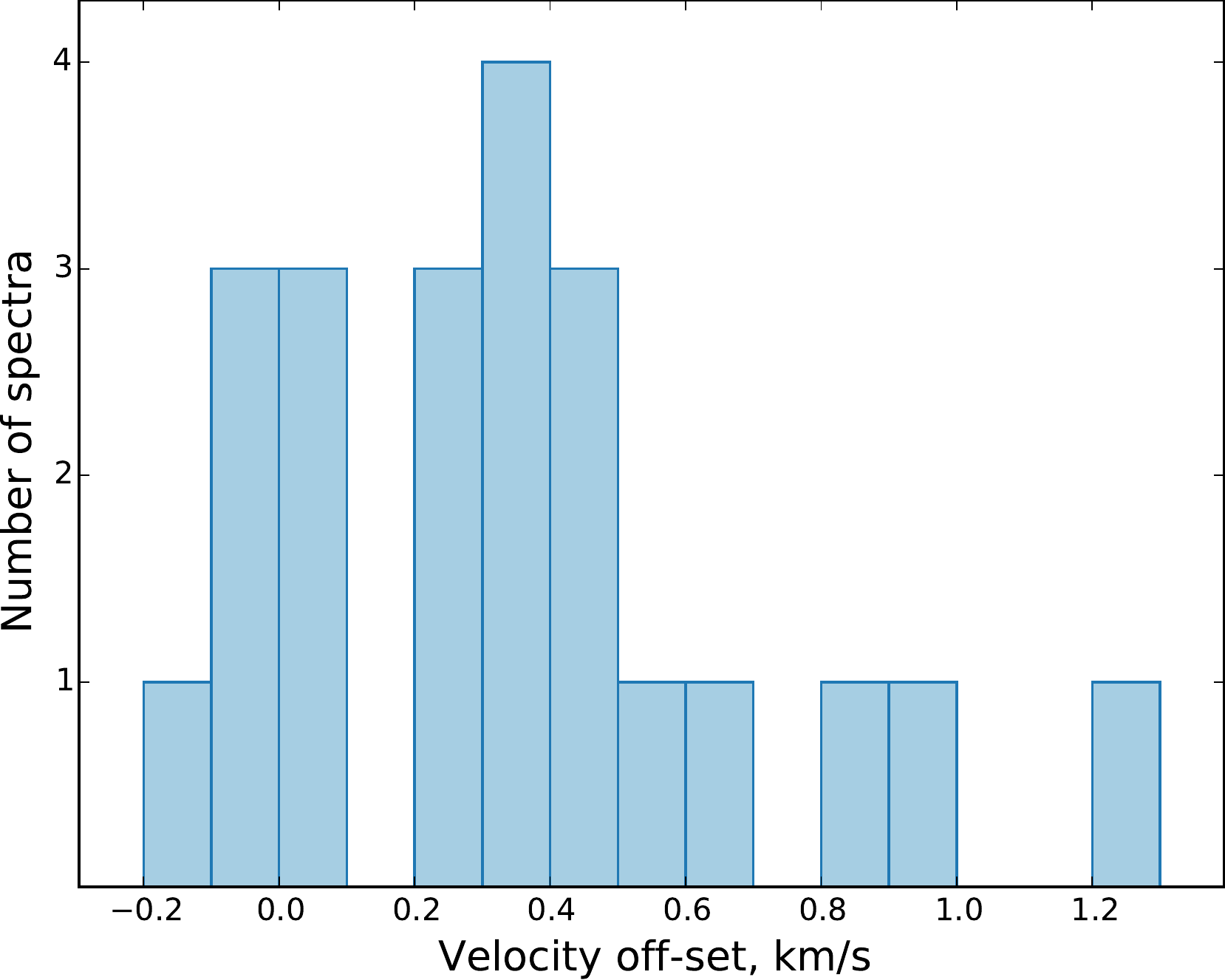}
\caption{Distribution of velocity offsets between all 22 exposures in the final sample. The offsets were calculated with respect to the exposure obtained on 29 March 2008 (starting time 06:40, UT).}
\label{fig:shift}
\end{figure}

In order to correct for possible velocity offsets between individual exposures, we use the following approach. Using purpose-written software\footnote{\url{https://github.com/ezavarygin/voffset}} we determine velocity offsets between the normalised one-dimensional spectra using a method similar to one described by \cite{Evans2013}. First, all exposures were convolved with a Gaussian (the assumed instrumental profile) whose values are given in Table~\ref{tab:j1011_journal}. Then flux and error arrays of one, anchor, exposure are modelled with cubic splines to get continuous functions of wavelength and the constant velocity offset, $\Delta v$: $f_{\rm o}(\lambda,\Delta v)$ and $\sigma_{\rm o}(\lambda,\Delta v)$ respectively. The velocity offset between the $j^{\rm th}$ exposure and the anchor exposure (subscript zero) is given by the minimum of the $\chi^2-$function:
\begin{equation}
\chi^2(\Delta v) = \sum_i\frac{[f_{\rm o}(\lambda, \Delta v) - f_{\rm j}(i)]^2}{\sigma_{\rm o}^2(\lambda,\Delta v) + \sigma_{\rm j}^2(i)},
\end{equation}
where $f_{\rm j}(i)$ and $\sigma_{\rm j}(i)$ are flux and error arrays, $i$ runs over all valid pixels unaffected by cosmic rays/CCD blemishes, not falling in gaps between the CCD chips and corresponding to a wavelength range covered by both the anchor and the examined spectrum. The wavelength in the anchor spectrum is calculated using a pixel-to-wavelength solution for the examined spectrum. The distribution of velocity offsets in our sample is illustrated in Fig.~\ref{fig:shift} showing that offsets between exposures up to 1.3~km~s$^{-1}$ appear to be present. 

Using this result, we co-add\footnote{A purpose-written python code was used: \url{https://github.com/ezavarygin/wspectrum}. } the 22 spectra, taking into account the offsets, into four spectra according to the decker and CCD binning in dispersion direction used at the exposures:
\begin{enumerate}
\item C1, unbinned (C1$\times$1)
\item C1, 2$\times$binned (C1$\times$2)
\item C5, unbinned (C5$\times$1)
\item C5, 2$\times$binned (C5$\times$2).
\end{enumerate}
Note, we also create a single co-added spectrum by combining the whole dataset. This spectrum is only used to demonstrate the overall quality of the data in Fig.~\ref{fig:data_quality} and to calculate an effective S/N around the transitions used (Table~\ref{tab:fitted_regions}). Apart from that, all final values and uncertainties stated in this work correspond to simultaneously fitting the four co-added spectra above.

\section{Analysis}

\subsection{Fitting}
\label{sec:fitting}

To fit the spectra we use the {\sc vpfit}\footnote{Version 10.2, \url{https://www.ast.cam.ac.uk/~rfc/vpfit.html}} software \citep{Carswell2014}. It uses a standard procedure of Voigt profile fitting to find a solution for redshift, $z$, column density, $N$, and the Doppler $b$ parameter (with an option to solve for both temperature, $T$, and the turbulent $b_{\rm turb}$ parameter) and their uncertainties for each of the absorbers.  {\sc vpfit} can also solve simultaneously for other parameters, including the local continuum and zero level.  {\sc vpfit} minimises a sum of the weighted squared residuals ($\chi^2$) between the model profile convolved with the instrumental resolution and the observed spectra, summed over the four co-added spectra.

During the fitting process, we adopt the following assumptions. For all the models, redshifts of H\,{\sc i} and D\,{\sc i} are tied in each component. The D/H ratio is assumed to be constant over the complex.  The metallicity of this system is  known to be low \citepalias{Burles1998b}, deuterium destruction is thus likely to be negligible, and the deuterium abundance should be very close to primordial  in this system.  In order to get a tighter constraint on the D/H ratio in the LLS, we solve for total H\,{\sc i} and D\,{\sc i} column densities rather than for column densities of the individual components. For models without metals we tie the Doppler parameters of H\,{\sc i} and D\,{\sc i} thermally (to be discussed in Section~\ref{sec:kin_structure}). When metals are used we solve for both $b_{\rm turb}$ and $T$. 

To take into account any possible residual velocity offsets between independent spectra, a velocity shift free-parameter is included in the model (the best-fit values lie close to zero: $\lesssim 0.1$~km~s$^{-1}$). For transitions with saturated lines, a zero level is fitted, allowing correction for a possibly imperfect night-sky subtraction during the data reduction process. 

\subsection{Continuum fitting}

Continuum fitting is a multi-step procedure.  During the spectral data reduction, the spectral orders from each individual echelle CCD exposure are extracted to one-dimensional spectra (see also Section~\ref{sec:j1011_reduction}). The orders from each individual exposure are placed into a one-dimensional array which is continuum fitted using polynomial fitting to unabsorbed spectral regions, and then normalised to unit continuum.  That process is repeated for each exposure/CCD so that multiple exposures can be combined using weighted addition to form a final co-added spectrum.

An example of the estimated continuum in the Ly\,$\alpha$ region for one of our 22 separate exposures is illustrated in the right hand panel of Fig.~\ref{fig:2_504_cont_norm}. The left hand panel of the same Figure illustrates the Lyman limit region from the same 3600 second exposure (corresponding to the last entry in Table \ref{tab:j1011_journal}).  Thus continuum fitting has already played an important role in the reduction process prior to analysis, during which further refinement of the continuum takes place.

Then, during VPFITting, each spectral segment is assigned additional free parameters to allow further adjustment of the continuum, using a linear fit, with free parameters slope and normalisation.  If there are insufficient continuum regions flanking the absorption line segment being fitted, only the normalisation is varied and the slope is assumed to be zero.  Importantly, the continuum parameters are adjusted simultaneously with all other model parameters. This means that the error estimates for the parameters of interest (in this case neutral hydrogen and deuterium column densities) properly take into account variation in the continuum parameters.

The reason for using a linear continuum fit simultaneous with profile modelling (rather than cubic or higher order) is that at the absorption profile modelling stage of the analysis, we are dealing with small wavelength segments (only a few \AA).   Non-linear continuum variations over such small scales are likely to be caused only by the presence of absorption lines and not by any other physical mechanism intrinsic to the quasar continuum.  If weak blends are found to be present, these are most appropriately taken into account by including additional absorption components.

Fig.~\ref{fig:ly_series} illustrates the fitting regions used as horizontal blue solid lines near the base of the absorption features.  Clearly estimating the continuum placement becomes less reliable as one approaches the Lyman limit.  The lower panel of Fig.~\ref{fig:ly_series} shows a region covering Lyman 21-24 where blending of high order Lyman lines results in there being no local absorption-free regions.  In this region, the extrapolated continuum was allowed to float but with no variation to the slope.

\begin{figure*}
\includegraphics[width=\linewidth]{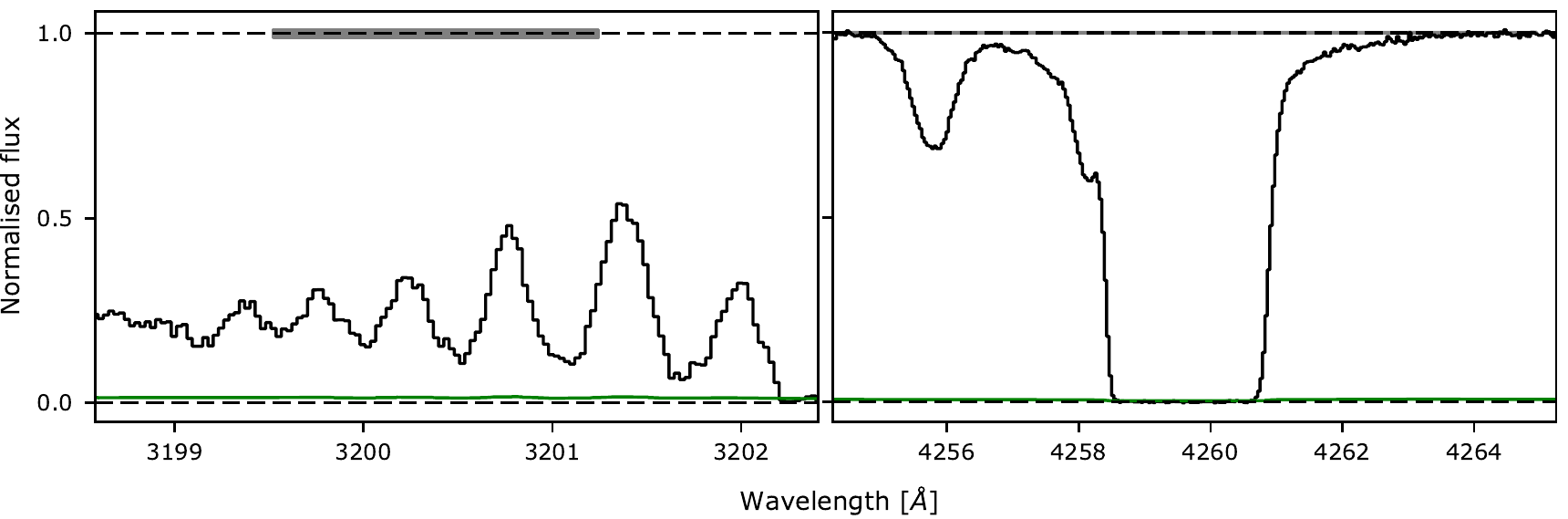}
\caption{A weighted spectrum of the $z_{\rm abs}=2.504$ LLS towards Q1009$+$2956, co-added from all available exposures. The left and right panels illustrate the Lyman Limit and the Ly\,$\alpha$ regions respectively. The black histogram is the flux, normalised to the best fit continuum corresponding to model 6a (see Sections~\ref{sec:kin_structure} and \ref{sec:2.504_multi_blending}). The grey shaded area near $y=1$ indicates the $1\sigma$ confidence region for the continuum fit. The green histogram near $y=0$ is the $1\sigma$ error in the normalised flux. The horizontal dashed lines indicate the zero and normalised continuum levels. The Figure illustrates the overall quality of the data and the continuum variation for model 6a.}
\label{fig:data_quality}
\end{figure*}

Fig.~\ref{fig:broad_abs} illustrates the final co-added spectra over the Ly\,$\alpha$ region for each of the four groupings of data described in Section \ref{sec:offset}.  The spectra are normalised to unit continuum, shown as horizontal black dashed lines.  The 1-$\sigma$ ranges of continuum parameters are illustrated as grey shaded areas.  The continua in all four cases are very well determined so the grey shaded area is small. The modelling was carried out simultaneously to the four spectral groupings.  However, Fig.~\ref{fig:data_quality} illustrates the overall data quality as the weighted co-addition of all four spectral groupings.  

\subsection{Lyman series}
\label{sec:ly_series}

\begin{figure*}
\center
\includegraphics[width=0.8\linewidth]{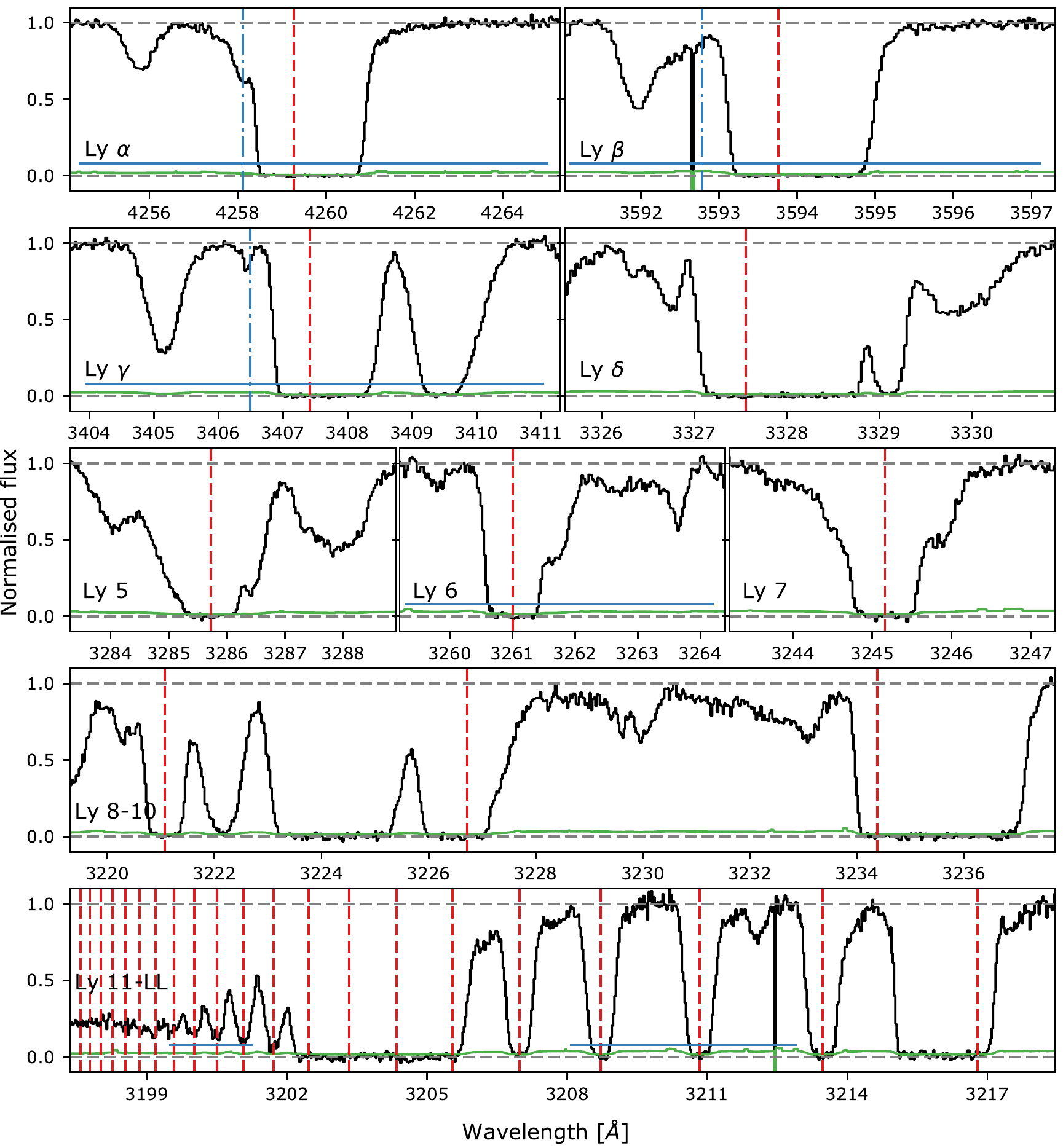}
\caption{Lyman series transitions for the $z_{\rm abs}=2.504$ Lyman limit system towards Q1009$+$2956 for the co-added C1$\times2$ spectrum. The green continuous line (around $y=0$) illustrates the 1$\sigma$ error array. The horizontal blue solid lines (at $y\simeq0.1$) indicate wavelength regions used in fitting (see Table~\ref{tab:fitted_regions}). The vertical red dashed lines indicate the center of the H\,{\sc i} Lyman series transitions in a one-component model (model 1a). The vertical blue dash-dotted line in the Ly\,$\alpha$, Ly\,$\beta$ and Ly\,$\gamma$ panels indicates the corresponding D\,{\sc i} position. The sharp drops of the flux around 3592.7 {\AA} in the top-right panel and 3212.5 {\AA} in the bottom panel are pixels clipped due to cosmic rays.}
\label{fig:ly_series}
\end{figure*}

All the H\,{\sc i} Lyman series transitions of the $z_{\rm abs}=2.504$ LLS down to the Lyman Limit have been covered by the exposures from our sample (Fig.~\ref{fig:ly_series}). The following transitions were included in the fitting process: Ly\,$\alpha$, Ly\,$\beta$, Ly\,$\gamma$, Ly\,6, Ly\,13, Ly\,14, Ly\,21-24 (Table~\ref{tab:fitted_regions}). Whilst other Lyman series lines were available, blending with lower redshift Lyman forest lines was more severe, such that those lines did not help to constrain the parameters of interest.  In addition to the main components of the LLS, there is one nearby component which has a fairly low H\,{\sc i} column density. It becomes optically thin at Ly\,5 and does therefore not show associated deuterium. This component lies $\sim +60$~km~s$^{-1}$ from the main absorption complex. It is fitted simultaneously with other parameters in order to properly take into account any impact it may have on determining the parameters of interest.

\begin{table}
\centering
\begin{minipage}{\linewidth}
\caption{Transitions and spectral ranges used in this work. }
\label{tab:fitted_regions}
\begin{tabular}{@{}lcccr@{}}
	\hline
	Transition  		   & Wavelength range$^a$, {\AA} & Spectra$^b$ & Cont.$^c$ & S/N$^d$ \\
	\hline
	Ly\,$\alpha$   	   & $4254.42 - 4265.00$ & $1 - 4$ & 2 & 147 \\
	Ly\,$\beta$ 		   & $3591.10 - 3597.10$ & $1 - 4$ & 2 & 99 \\
	Ly\,$\gamma$        & $3403.95 - 3411.03$ & $1 - 4$ & 1 & 92  \\	
	Ly\,6               & $3259.30 - 3264.20$ & $1 - 3$ & 1 & 51 \\
	Ly\,13    		   & $3209.81 - 3212.89$ & $1 - 4$ & 1 & 51 \\
	Ly\,14		       & $3208.08 - 3210.20$ & $1 - 4$ & 1 & 50 \\
	Ly\,21-24   		   & $3199.50 - 3201.25$ & $1 - 4$ & 1 & 27 \\
	Si\,{\sc iv} 1393  & $4882.30 - 4884.10$ & $1 - 4$ & 2 & 128 \\
	Si\,{\sc iv} 1402  & $4913.90 - 4915.70$ & $1 - 4$ & 2 & 140 \\
	C\,{\sc ii} 1334   & $4674.80 - 4676.70$ & $1 - 4$ & 2 & 106 \\
	C\,{\sc iii} 977   & $3422.22 - 3425.60$ & $1 - 4$ & 1 & 80 \\
	C\,{\sc iv} 1548   & $5423.20 - 5425.50$ & $1 - 4$ & 2 & 119 \\
	C\,{\sc iv} 1550   & $5432.25 - 5434.60$ & $1 - 3$ & 2$^e$ & 98 \\
	\hline
\end{tabular}

\small{$^a$Wavelength ranges used to fit the transition.}\\
\small{$^b$Co-added spectra used in fitting out of 4 listed in Section~\ref{sec:offset}. The 4th spectrum, C$5\times2$, was excluded for Ly\,6 and C\,{\sc iv} 1550 due to very low S/N and/or cosmic ray contamination.}\\
\small{$^c$Order of polynomials used to fit the continuum in the region: 1 -- floating continuum, 2 -- floating continuum with a non-zero slope.}\\
\small{$^d$Average signal-to-noise ratio measured at continuum level around the transition. For the Ly\,21-24 transitions the least absorbed pixels are used since the flux does not return to the continuum level.}\\
\small{$^e$Except the C5$\times$1 spectrum where the slope is not allowed to vary due to cosmic ray contamination.}\\
\end{minipage}
\end{table}

An apparent deuterium feature is clearly seen at its expected position ($\sim -81.5$~km~s$^{-1}$ bluewards from the centre of the H\,{\sc i} line) of Ly\,$\alpha$ at $z_{\rm abs}\approx2.5036$. Due to blending with low-redshift Ly\,$\alpha$ forest lines the deuterium absorption is not visible in higher order Lyman series transitions.

\subsection{Metals}
\label{sec:2_504_metals}

Given line saturation, blending, and finite S/N, it is unlikely that one could reliably determine the velocity structure of the LLS based on the Lyman series alone. A more reliable estimate of velocity structure is obtained by simultaneously fitting both Lyman series lines and metal species.
\begin{figure}
\center
\includegraphics[width=\linewidth]{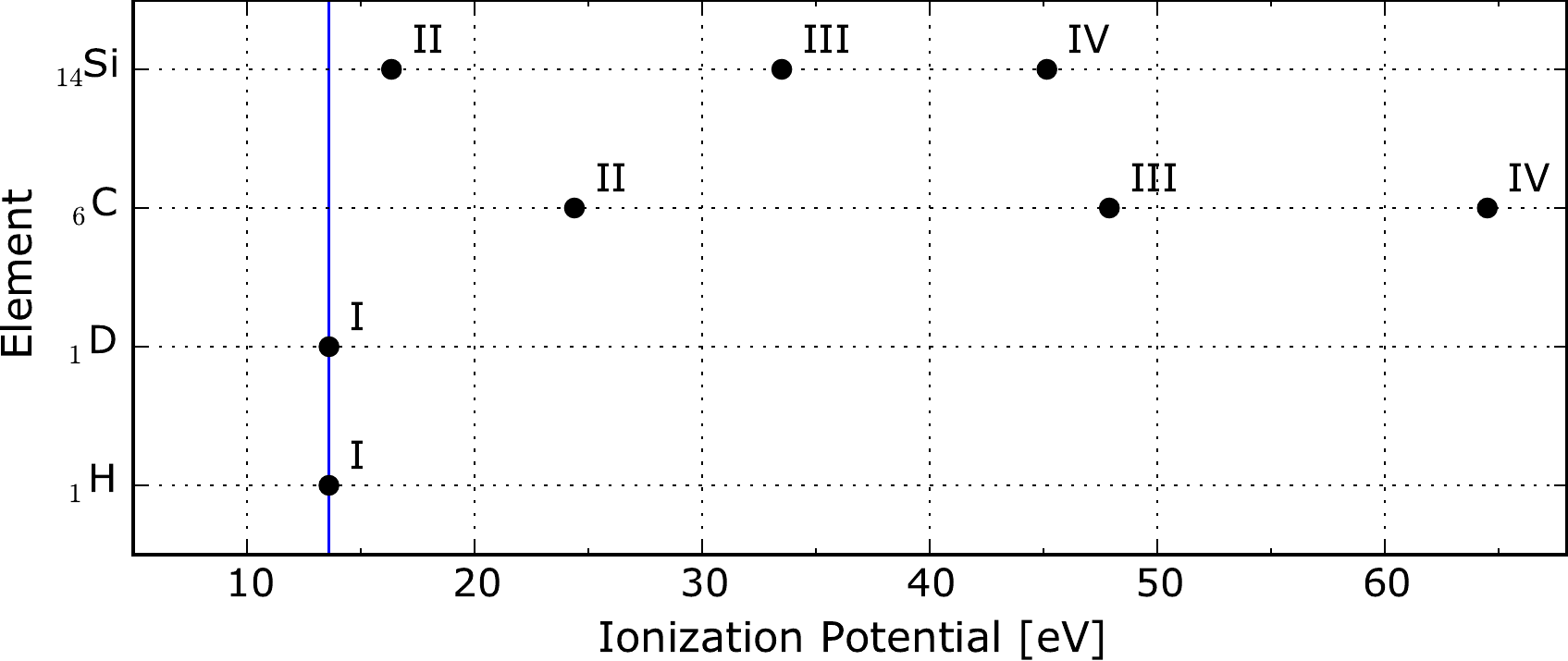}
\caption{Ionisation potential values for the transitions detected in the $z_{\rm abs}=2.504$ Lyman limit absorption system. The vertical solid line indicates the ionisation potential of H\,{\sc i}, 13.5984~eV. The values were taken from the NIST Atomic Spectra Database \citep{Kramida2017}.}
\label{fig:ionization_plot}
\end{figure}

The LLS at $z_{\rm abs}=2.504$ shows absorption by carbon and silicon from different ionisation levels: C\,{\sc ii}, C\,{\sc iii}, C\,{\sc iv}, Si\,{\sc ii}, Si\,{\sc iii} and Si\,{\sc iv}. Fig.~\ref{fig:ionization_plot} illustrates the relevant ionisation potentials. Absorption lines from C\,{\sc ii} 1334, C\,{\sc iii} 977, C\,{\sc iv} 1548 and 1550, Si\,{\sc ii} 1260, Si\,{\sc iii} 1206, Si\,{\sc iv} 1393 and 1402 are detected. 
C\,{\sc iii}, C\,{\sc iv} and Si\,{\sc iv} are all strong. Fitting each of these features independently shows consistent structures with at least two strong components. However, a 2-component simultaneous fit to C\,{\sc iii}, C\,{\sc iv} and Si\,{\sc iv} with tied redshifts and b-parameters does not yield a statistically acceptable fit.  A 3-component model does however fit well preferring a thermal broadening over the turbulent one (see Section~\ref{sec:kin_structure}). C\,{\sc ii} is also consistent with C\,{\sc iii}, C\,{\sc iv} and Si\,{\sc iv} but is weak so provides little additional constraint.  Fig.~\ref{fig:2_504_model_6} illustrates the transitions used. The Si\,{\sc iii} 1206~{\AA} line was not used due to a problem with the rest frame wavelength (see Section \ref{sec:SiIII}). Si\,{\sc ii} 1260 is very weak and is also badly blended so was discarded. Table~\ref{tab:fitted_regions} contains information about the transitions and wavelength ranges used to constrain the parameters of interest.

\subsubsection{Si {\sc iii}}
\label{sec:SiIII}

\cite{Morton2003} provides a Si\,{\sc iii}~1206 rest-frame wavelength of 1206.500\,{\AA}. Using this wavelength leads to an apparent shift of Si\,{\sc iii}~1206 redwards with respect to C\,{\sc iii}, C\,{\sc iv}, and Si\,{\sc iv} by $\sim+2.5$~km~s$^{-1}$. The NIST spectral database gives an observed wavelength of 1206.51\,{\AA} \citep{Kramida2017}. The 0.01\,{\AA} difference between the Morton and NIST wavelength corresponds to 2.49\,km~s$^{-1}$ in the same direction. We conclude that the most likely explanation of the observed shift between Si\,{\sc iii} and other transitions is simply due to the fact that the true Si\,{\sc iii} 1206 wavelength is closer to the old experimental value of 1206.51\,{\AA} than the value listed in Morton (1206.500\,{\AA}). Due to this uncertainty, to avoid any possible bias we exclude Si\,{\sc iii}~1206 from our analysis -- new experimental data for this transition is clearly needed.

Another potential explanation of the shift, in case the Morton value is correct, is that the Si\,{\sc iii} 1206\,{\AA} line is blended with some unidentified interloper.  If so, this is unlikely to be hydrogen (even though the redshifted Si\,{\sc iii} 1206 line falls in the forest) because the observed line width is far smaller than typical forest $b-$parameters. We searched for potential interlopers at the redshifts of other absorption systems identified along the line of sight (and from the interstellar medium of our own galaxy) but did not find any likely candidates (see Section~\ref{sec:poss_cont}).

\subsection{Possible contamination}
\label{sec:poss_cont}

There is always a chance that a transition of interest is blended with absorption from another system at a different redshift. It is therefore important to try and identify all absorption systems present in the spectrum.  This was done in two ways.  Firstly, a cross-correlation method\footnote{\url{https://github.com/TrisD/Ab_Detect}} was used to compare a list of standard metal lines with the list of the absorption lines detected in the spectrum.  Secondly, a visual inspection of the spectrum was carried out, scanning in redshift, using a template of standard lines, to identify coincidences in velocity space for transitions in the same system.  The whole process resulted in the identification of 24 absorption systems (including the $z_{\rm abs}=2.504$ system of interest) given in Appendix~\ref{app:abs_systems}.  We then check whether any commonly detected transitions at each candidate redshift fall within $\pm 100$~km~s$^{-1}$ of any transition in our $z_{\rm abs}=2.504$ system of interest\footnote{The following routine has been used: \url{https://github.com/vincentdumont/zblend}.}.  Nevertheless, despite a careful search for potential contamination of this sort, none was found.

\subsection{Fitting different models}
\label{sec:kin_structure}

Weak absorption lines due to metals with low ionisation potentials (Si\,{\sc ii}, C\,{\sc ii}) are detected. We do not detect any neutral species, e.g. O\,{\sc i}. The relative abundances of the various ionisation stages, can vary significantly for each velocity component across the absorbing medium.  Components detected in Si\,{\sc iv} and C\,{\sc iv} for example, may have weak (even undetectable) H\,{\sc i}.  This is of course taken into account in the modelling procedure because all column densities are free parameters. 

\citetalias{Burles1998b} concluded that the velocity dispersions of the H\,{\sc i} and D\,{\sc i} lines are dominated by thermal motions. We agree with this finding. 
In all models using metal transitions that we consider (see below), we fitted each of the main absorption components with additional free parameters corresponding to temperature and any possible turbulent component of the line broadening, $b_{\rm turb}$.
For all components, $b_{\rm turb}$ is small compared to the observed $b$-parameter ($<5\%$ for H\,{\sc i}), given by $b_{\rm obs}^2 = b_{\rm therm}^2 + b_{\rm turb}^2$.  The full set of model parameters are provided as online MNRAS supplementary files and on GitHub (see links in Section~\ref{sec:2_504_fit_results}).  As a further check on this point, we fit metals both thermally and turbulently with a three-component model. A thermal fit is statistically preferable with $\chi^2/{\rm dof}=0.678$ vs 0.763 for the turbulent fit. For models in which only the Lyman series is fitted (i.e. without simultaneously fitting to metal lines; see below), we therefore assume thermal broadening, thus avoiding degeneracy between $T$ and $b_{\rm turb}$.

In order to explore the possible impact on D/H of different assumptions about the velocity structure, we explore six different models for the absorption complex, where the H\,{\sc i} components are tied (in redshift and $b$-parameters) to different combinations of metal lines:
\begin{enumerate}
\item[1)] one-component model, fit to Lyman series only;
\item[2)] two-component model, fit to Lyman series only;
\item[3)] three-component model, fit to Lyman series only;
\item[4)] two-component model, C\,{\sc ii} used;
\item[5)] three-component model, C\,{\sc ii} used;
\item[6)] three-component model, C\,{\sc ii}, C\,{\sc iii}, C\,{\sc iv} and Si\,{\sc iv} used.
\end{enumerate}

We assume throughout that D/H is the same across all components. The first three models use Lyman series transitions only. Since metals strongly support a thermal broadening, in these three models we tie H\,{\sc i} and D\,{\sc i} thermally. Even though model 1 is clearly unrealistic (at least two strong components are seen in the low ionisation metal species), we include it for completeness and note that despite being too simple, it gives a reasonably consistent result for D/H (Section~\ref{sec:2_504_fit_results}). Models 4 and 5 exclude the higher ionisation species, including only C\,{\sc ii}~1334 as this line appears to be unblended. Model 6 is a simultaneous fit to four metal species plus the hydrogen and deuterium series.  The parameters ($z$, $b_{\rm turb}$ and $T$) of H\,{\sc i} and D\,{\sc i} were tied with those of metals in models 4 through 6.

\begin{figure*}
\center
\includegraphics[width=0.9\linewidth]{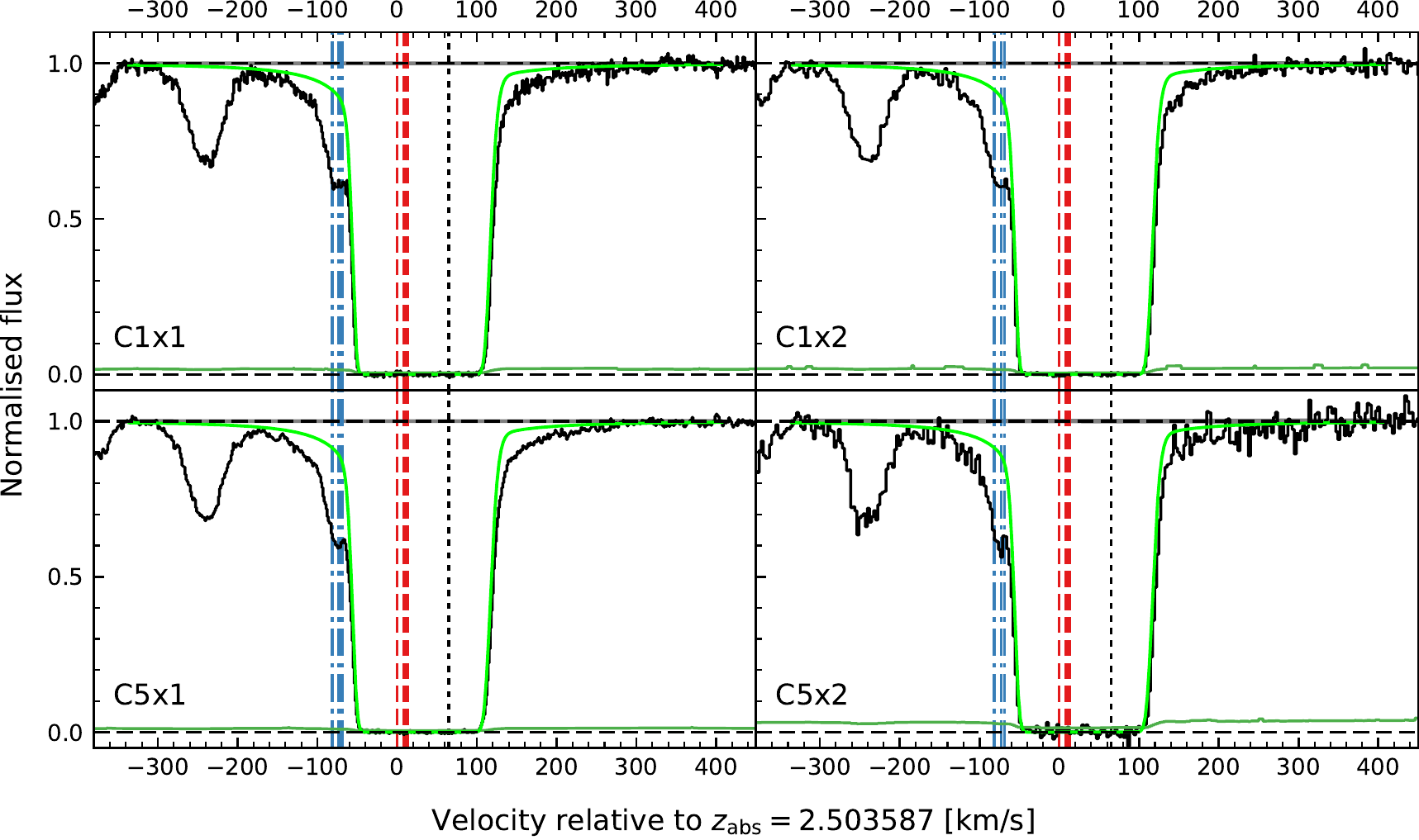}
\caption{The Ly\,$\alpha$ transition for the $z_{\rm abs}=2.504$ Lyman limit absorption system for each of the four independent datasets, as indicated at the bottom left corner of each panel. The continuous green line is formed from the three H\,{\sc i} components in model 6a (vertical red dashed lines) that give rise to the D\,{\sc i} absorption feature (their expected position is indicated with vertical blue dash-dotted lines), plus one additional H\,{\sc i} component (vertical black dotted line) that is required  to account for the  additional absorption redwards of the three main components.  The 1-$\sigma$ error array is shown by green histogram close to $y=0$. The grey (very narrow) shaded area near $y=1$ corresponds to the $\pm1\sigma$ uncertainty in the continuum. The purpose of this Figure is to illustrate the additional broad absorption wings either side of the main, strong, H\,{\sc i} components.}
\label{fig:broad_abs}
\end{figure*}

The  total H\,{\sc i} column density is too low to explain the Lorentzian-like wings that are particularly visible in the wings of Ly\,$\alpha$ (see Fig.~\ref{fig:broad_abs}).  These broad wings appear to be symmetric with respect to the profile associated only with the sharp central absorption feature.  We therefore explore whether the broad residual absorption is most likely to be explained by multiple narrow blends in the Ly\,$\alpha$ wings or whether one large b-parameter, spread across the whole profile, provides the ``correct'' model. These two groups of models are indicated as 1a$-$6a and 1b$-$6b, respectively.

\subsubsection{Modelling the broad wings in H\,{\sc i} Ly\,$\alpha$ using multiple narrow blends}
\label{sec:2.504_multi_blending}

\begin{figure*}
\center
\includegraphics[width=\textwidth]{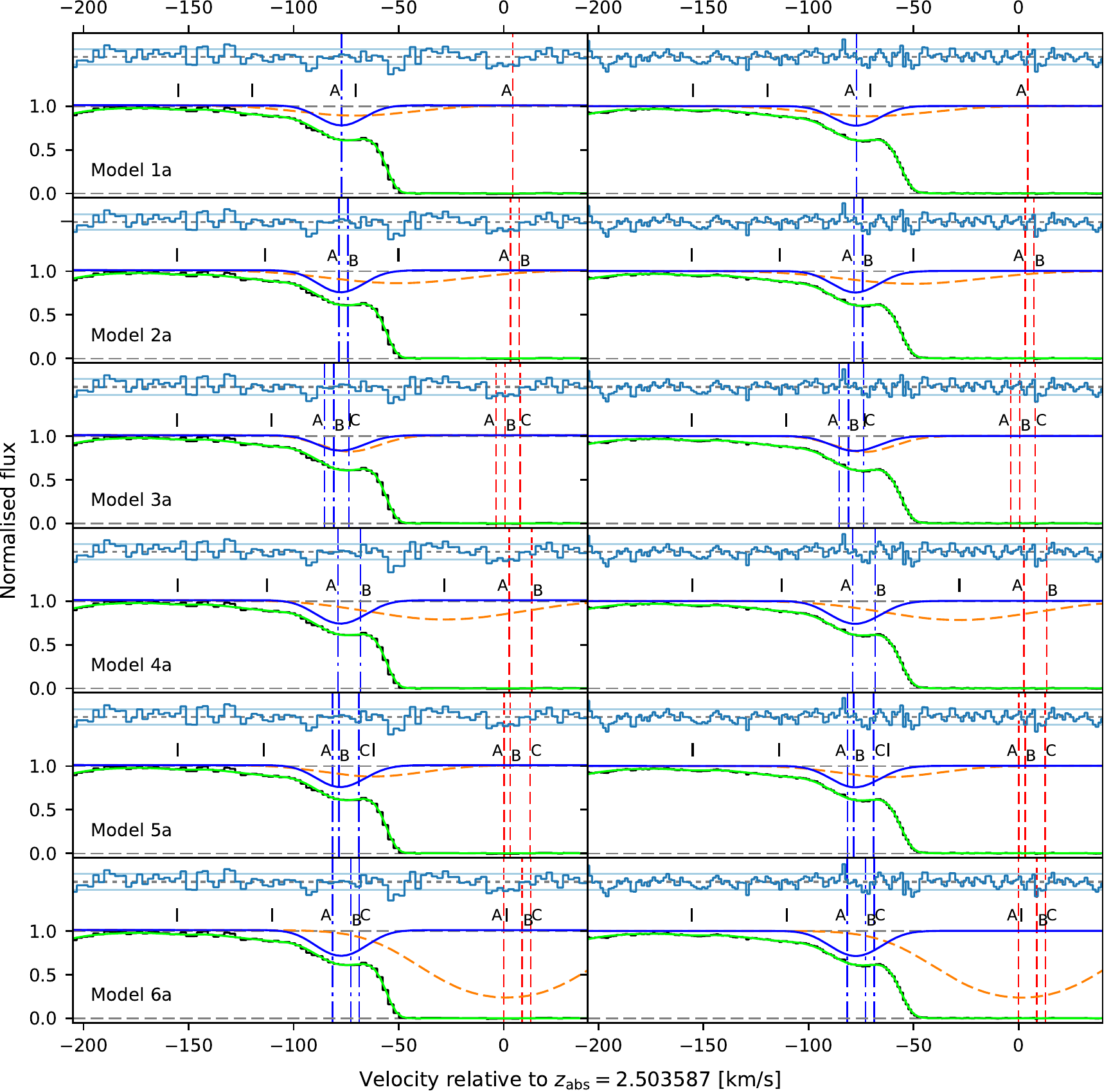}
\caption{The overall absorption profile (green solid line) for the blue wing of the H\,{\sc i} Ly\,$\alpha$ transition for the six considered models using multiple narrow blends (model 1a is at the top) described in Section~\ref{sec:2.504_multi_blending}. Left and right panels correspond to the C$1\times2$ and C$5\times1$ spectra (black histogram) respectively. These two co-added spectra illustrated are the highest S/N. However all four co-added spectra were fitted. The residuals with $1\sigma$ confidence range are shown in the blue histogram above each fit. The main absorption components are indicated with vertical red dashed lines with letters above. The positions of the corresponding deuterium transitions are indicated with the blue vertical dash-dotted lines. The ticks above each fit indicate absorption by additional H\,{\sc i} clouds included in the model. The blue solid line illustrates the overall D\,{\sc i} absorption.  The orange dashed line shows the additional contaminating H\,{\sc i} line that degrades the measurement precision on D\,{\sc i}. The x-axis show the velocity shift relative to component A for model 6a. }
\label{fig:2.504_blend_profile}
\end{figure*}

We first assume that the broad wings are caused by multiple weak, narrow blends.  The severe blending means that the line parameters for the multiple weak lines needed to model the wings adequately are poorly constrained. The blends in the red wing do not affect the D/H estimate substantially since the impact of these weak blends on $\log N$(H\,{\sc i}) for the stronger central absorption is negligible.  However, the broad residual absorption in the blue wing of Ly\,$\alpha$ can also be modelled using several narrow blends, one of which lies close to the D feature and which, unfortunately, significantly degrades the precision with which we can determine D/H for this system.  Fig.~\ref{fig:2.504_blend_profile} illustrates the details in the blue wing of H\,{\sc i} Ly\,$\alpha$ for models 1a$-$6a.

The S/N of our spectra is much higher than the spectrum studied by \citetalias{Burles1998b} ($\sim$147 vs 60 in the continuum level around Ly\,$\alpha$). \citetalias{Burles1998b} suggested that the deuterium line is blended with a Ly\,$\alpha$ forest line at approximately $-30$~km~s$^{-1}$ of the the D\,{\sc i} Ly\,$\alpha$ line (see their fig.~8a). The \citetalias{Burles1998b} model contains only this one blend in the vicinity of the D line.  Using narrow lines, we found a best-fit to the blue wing of Ly\,$\alpha$ using three components, illustrated in Fig.~\ref{fig:2.504_blend_profile}, which shows all of our six models.

We can estimate the impact of this blend on the D/H ratio precision by fixing the parameters of the blend and running {\sc vpfit} once again. For example, for three-component models 3a, 5a and 6a, fixing the parameters of the blend leads to significantly reduced D/H uncertainties: from 0.5 to 0.027~dex, from 0.25 to 0.018~dex and from 0.08 to 0.014~dex, respectively. 

The blend parameters and therefore D/H are sensitive to the initial continuum level (Fig.~\ref{fig:2_504_cont_norm}). We include a floating continuum with a varying slope over the Ly\,$\alpha$ region in our models, allowing for fine-tuning of the continuum placement.  The apparent blend close to the D line is not explained by a simple fluctuation in continuum fitting since it would require a $10-15\%$ variation (see Fig.~\ref{fig:2.504_blend_profile}) over the D\,{\sc i} Ly\,$\alpha$ line.  Such a strong variation over such a small scale (less than 2~{\AA} or 140\,km\,s$^{-1}$) is unlikely.

We raise one further point concerning the apparent blending of the D\,{\sc i}.  We opted to interpret the `extra' absorption as being due to an interloping H\,{\sc i} line.  The data are consistent with this interpretation and indeed this is the most likely scenario.  If we had instead interpreted the residual absorption as being due to D\,{\sc i}, a corresponding additional H\,{\sc i} component would obviously be required.  If we maintain the assumption of constant D/H in all velocity components, the H\,{\sc i} column density in this additional component would be too high (higher order Lyman lines are unsaturated at the corresponding velocity), thereby ruling out that interpretation.  On the other hand, if we were to relax the assumption of constant D/H for all velocity components, the additional component in question would then permit a very high D/H in that specific component, requiring an inhomogeneous BBN explanation or subsequent inhomogeneous astration.  Whilst we cannot rule that scenario out, we have not explored it in further detail and simply acknowledge this potential bias in interpreting the data.

\subsubsection{Modelling the broad wings in H\,{\sc i} Ly\,$\alpha$ using one single broad line}
\label{sec:2.504_broad_blending}

\begin{figure*}
\center
\includegraphics[width=\textwidth]{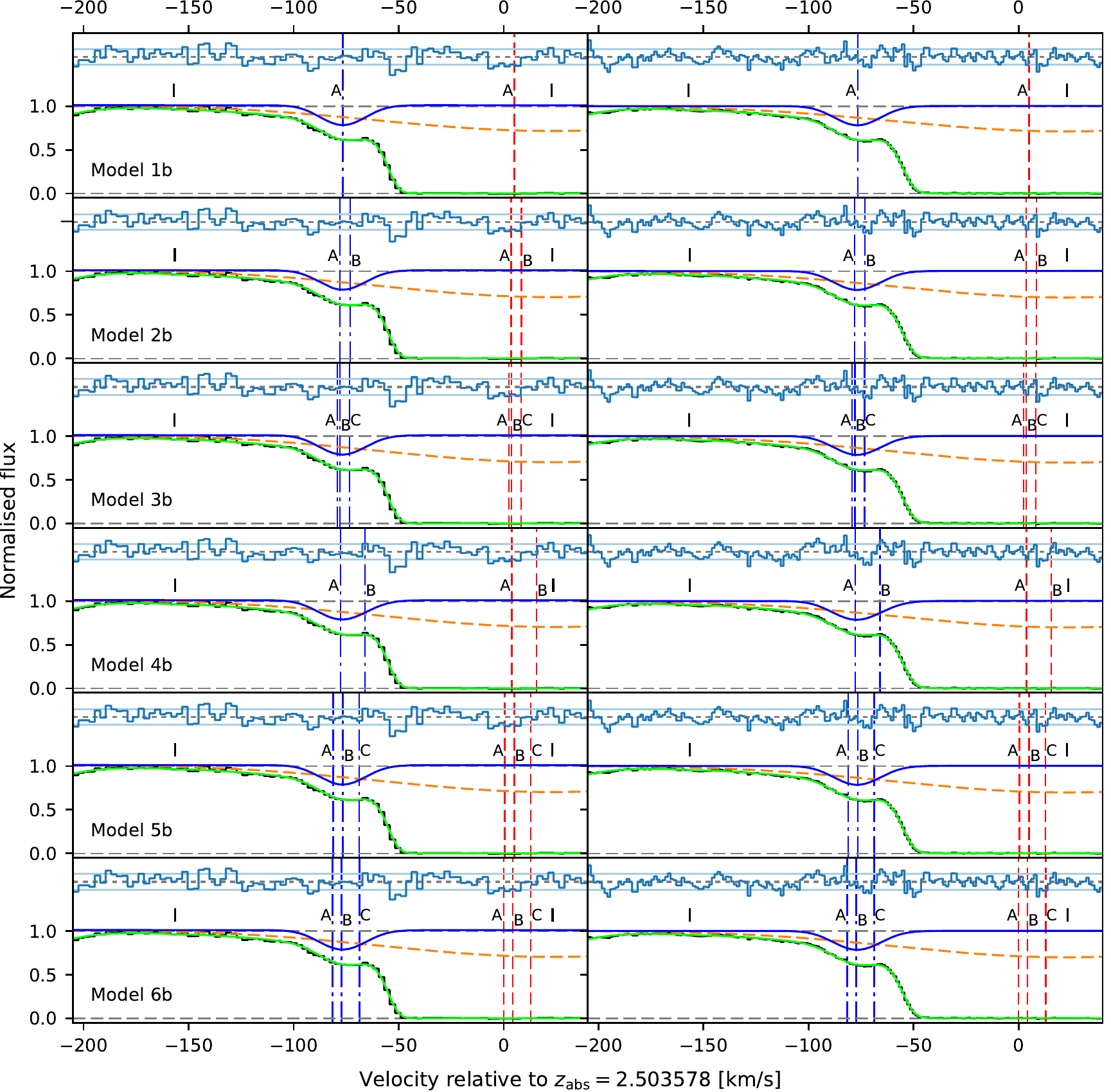}
\caption{Same as Fig.~\ref{fig:2.504_blend_profile} but for models 1b$-$6b using a single broad line (shown with the orange dashed line) to fit broad wings in H\,{\sc i} Ly\,$\alpha$.}
\label{fig:2.504_blend_broad_profile}
\end{figure*}

Given the symmetry of the additional absorption in the wings of the Ly\,$\alpha$ line, instead of using multiple narrow blends to accurately model the broad wings, we find that a single broad ($b\sim105$\,km\,s$^{-1}$) line sitting on top of Ly\,$\alpha$ also provides a good fit to the data. In doing so, the parameters of the single broad blend are very well constrained, and the model yields an extremely high D/H precision of (see Section~\ref{sec:2_504_fit_results}).  Fig.~\ref{fig:2.504_blend_broad_profile} illustrates the details in the blue wing of H\,{\sc i} Ly\,$\alpha$ for models 1b$-$6b.

\subsection{Fitting results}
\label{sec:2_504_fit_results}

\begin{figure}
\center
\includegraphics[width=\linewidth]{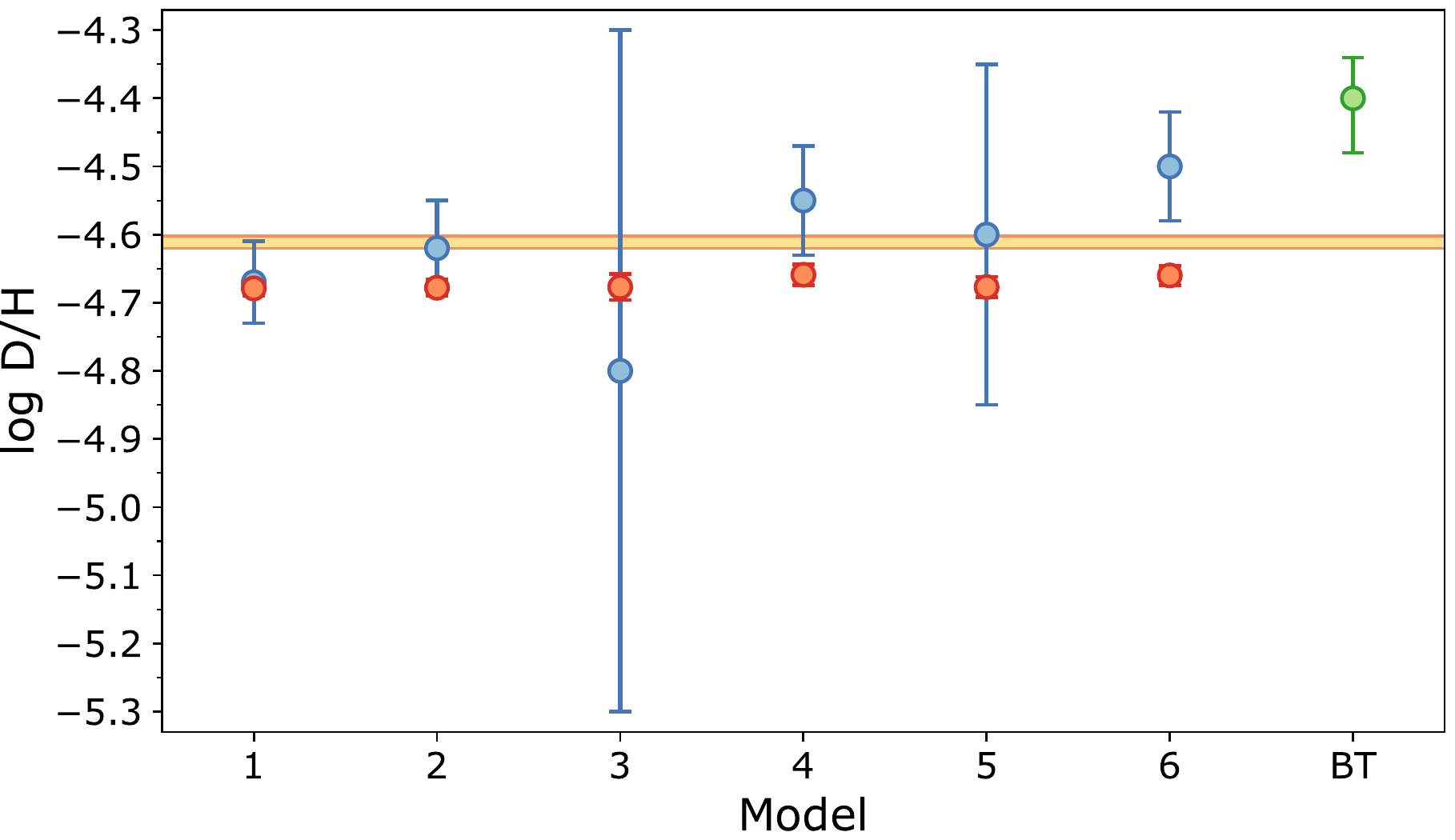}
\caption{D/H values with 1$\sigma$ confidence intervals (y-axis) for each of the six pairs of models considered (indicated on the x-axis). The blue points (larger error bars) correspond to models 1a$-$6a.  The orange points (smaller error bars) correspond to models 1b$-$6b (see Sections~\ref{sec:2.504_multi_blending} and \ref{sec:2.504_broad_blending}). The \citetalias{Burles1998b} measurement is shown at the far right. The horizontal band indicates the 1$\sigma$ confidence interval for the CMB value from \protect\cite{Coc2015}.}
\label{fig:d2h_results}
\end{figure}

Best-fit values of the total H\,{\sc i} and D\,{\sc i} column densities and corresponding D/H ratios with 1$\sigma$ confidence intervals for each of the models considered are presented in Tables~\ref{tab:d2h_results} and \ref{tab:d2h_results_broad}. The total H\,{\sc i} column density is very well constrained, is consistent amongst all our models and agrees with the constraints from \citetalias{Burles1998b}. Fig.~\ref{fig:d2h_results} illustrates the D/H values for our models. 

\begin{table*}
\centering
\caption{The best-fit D/H values for the models described in Section~\ref{sec:kin_structure} using multiple narrow blends to fit the broad H\,{\sc i} Ly\,$\alpha$ wings, as described in Section~\ref{sec:2.504_multi_blending}. All the {\sc vpfit} files with the best-fit parameters and their uncertainties for each of the models are available in online supplementary files, given in the links in Section~\ref{sec:2_504_fit_results}.}
\label{tab:d2h_results}
\begin{tabular}{@{}cclccccll@{}}
\hline
Model & N$^{\underline{\rm o}}$ components & Metals used  & $\chi^2$/dof & AICc & $\Delta$AICc$^a$ & log~$N_{\rm total}$(H\,{\sc i}) & log~$N_{\rm total}$(D\,{\sc i}) & log~D/H \\ 
\hline
1a     &	1             & ---   & 0.71911 & 4993.7 & 18.1 & $17.365 \pm 0.005$ & $12.70 \pm 0.06$ & $-4.67 \pm 0.06$ \\
2a     &	2             & ---   & 0.71636 & 4980.2 & 4.6 & $17.363 \pm 0.005$ & $12.75 \pm 0.07$ & $-4.62 \pm 0.07$ \\
3a     &	3             & ---   & 0.71500 & 4975.6 & 0 & $17.363 \pm 0.005$ & $12.6 \pm 0.5$   & $-4.8 \pm 0.5$   \\
4a     &	2             & C\,{\sc ii} & 0.71400 & 5180.5 & 7.2 & $17.360 \pm 0.005$ & $12.81 \pm 0.08$ & $-4.55 \pm 0.08$ \\
5a     &	3             & C\,{\sc ii} & 0.71190 & 5173.3 & 0 & $17.363 \pm 0.005$ & $12.77 \pm 0.25$ & $-4.60 \pm 0.25$ \\
6a     &	3             & C\,{\sc ii}, C\,{\sc iii}, C\,{\sc iv}, Si\,{\sc iv} & 0.71520 & 6492.6 & 0 & $17.360 \pm 0.005$ & $12.86 \pm 0.08$ & $-4.50 \pm 0.08$ \\
\hline
\multicolumn{8}{l}{CMB prediction by \cite{Coc2015}} & $-4.611\pm0.009$ \\
\hline
\end{tabular}

\small{$^a\Delta\text{AICc} = \text{AICc} - \text{AICc}_{\rm min}$, where AICc$_{\rm min}$ corresponds to the model with the smallest AICc within each of the three subsets of models (no metals, C\,{\sc ii}, and all metals).}
\end{table*}

\begin{table*}
\centering
\caption{Same as in Table~\ref{tab:d2h_results} but with a single high b-parameter line used to model the broad wings of H\,{\sc i} Ly\,$\alpha$, as described in Section~\ref{sec:2.504_broad_blending}.}
\label{tab:d2h_results_broad}
\begin{tabular}{@{}cclccccll@{}}
\hline
Model & N$^{\underline{\rm o}}$ components & Metals used  & $\chi^2$/dof & AICc & $\Delta$AICc$^a$ & log~$N_{\rm total}$(H\,{\sc i}) & log~$N_{\rm total}$(D\,{\sc i}) & log~D/H \\ 
\hline
1b     &	1             & ---   & 0.72705 & 5032.6 & 5.5 & $17.367 \pm 0.005$ & $12.688 \pm 0.010$ & $-4.679 \pm 0.011$ \\
2b     &	2             & ---   & 0.72556 & 5027.1 & 0   & $17.363 \pm 0.005$ & $12.685 \pm 0.011$ & $-4.678 \pm 0.012$ \\
3b     &	3             & ---   & 0.72592 & 5033.6 & 6.5 & $17.363 \pm 0.005$ & $12.687 \pm 0.019$ & $-4.677 \pm 0.019$ \\
4b     &	2             & C\,{\sc ii} & 0.72344 & 5231.4 & 9.8 & $17.363 \pm 0.005$ & $12.704 \pm 0.014$ & $-4.659 \pm 0.015$ \\
5b     &	3             & C\,{\sc ii} & 0.72095 & 5221.6 & 0 & $17.364 \pm 0.005$ & $12.687 \pm 0.014$ & $-4.677 \pm 0.015$ \\
6b     &	3             & C\,{\sc ii}, C\,{\sc iii}, C\,{\sc iv}, Si\,{\sc iv} & 0.72228 & 6539.3 & 0 & $17.360 \pm 0.005$ & $12.700 \pm 0.013$ & $-4.660 \pm 0.014$ \\
\hline
\multicolumn{8}{l}{CMB prediction by \cite{Coc2015}} & $-4.611\pm0.009$ \\
\hline
\end{tabular}

\small{$^a$Calculated with respect to the models with a broad line fitted.}
\end{table*}

Model 6a is illustrated in Figures~\ref{fig:2_504_model_6} (a) and (b) and provides a velocity structure comparison between hydrogen and metal lines. 
The data plotted are C1$\times$2 and C5$\times$1 respectively.  These two co-added spectra are the highest S/N. Models 1a$-$5a give similar-looking residuals (see Fig.~\ref{fig:2.504_blend_profile}). The blue wing of Ly\,$\alpha$ for models 1b$-$6b is shown in Fig.~\ref{fig:2.504_blend_broad_profile}.

For model 6a, we demonstrate the justification for including the Ly\,21-24 region in the modelling procedure.  If we exclude the region entirely, and re-fit, we obtain log\,N$_{\rm total}$(H\,{\sc i})~$=17.355\pm0.014$.  This compares to log\,N$_{\rm total}$(H\,{\sc i})~$=17.360\pm0.005$ when the region is included in the fit (with a local freely varying continuum). We thus see a factor of almost 3 improvement in the precision of the total neutral hydrogen column density measurement.  The actual column densities in both cases are consistent and D/H does not alter significantly.

All {\sc vpfit} output files with the parameters and uncertainties for all the models considered are available in online supplementary files\footnote{MNRAS supplementary files url.} and also on GitHub\footnote{\url{https://github.com/ezavarygin/q1009p2956}}. A short description of the supplementary files is given in Appendix~\ref{app:suppl_files}.

\begin{figure*}
\begin{subfigure}{\linewidth}
\centering
\includegraphics[width=\linewidth]{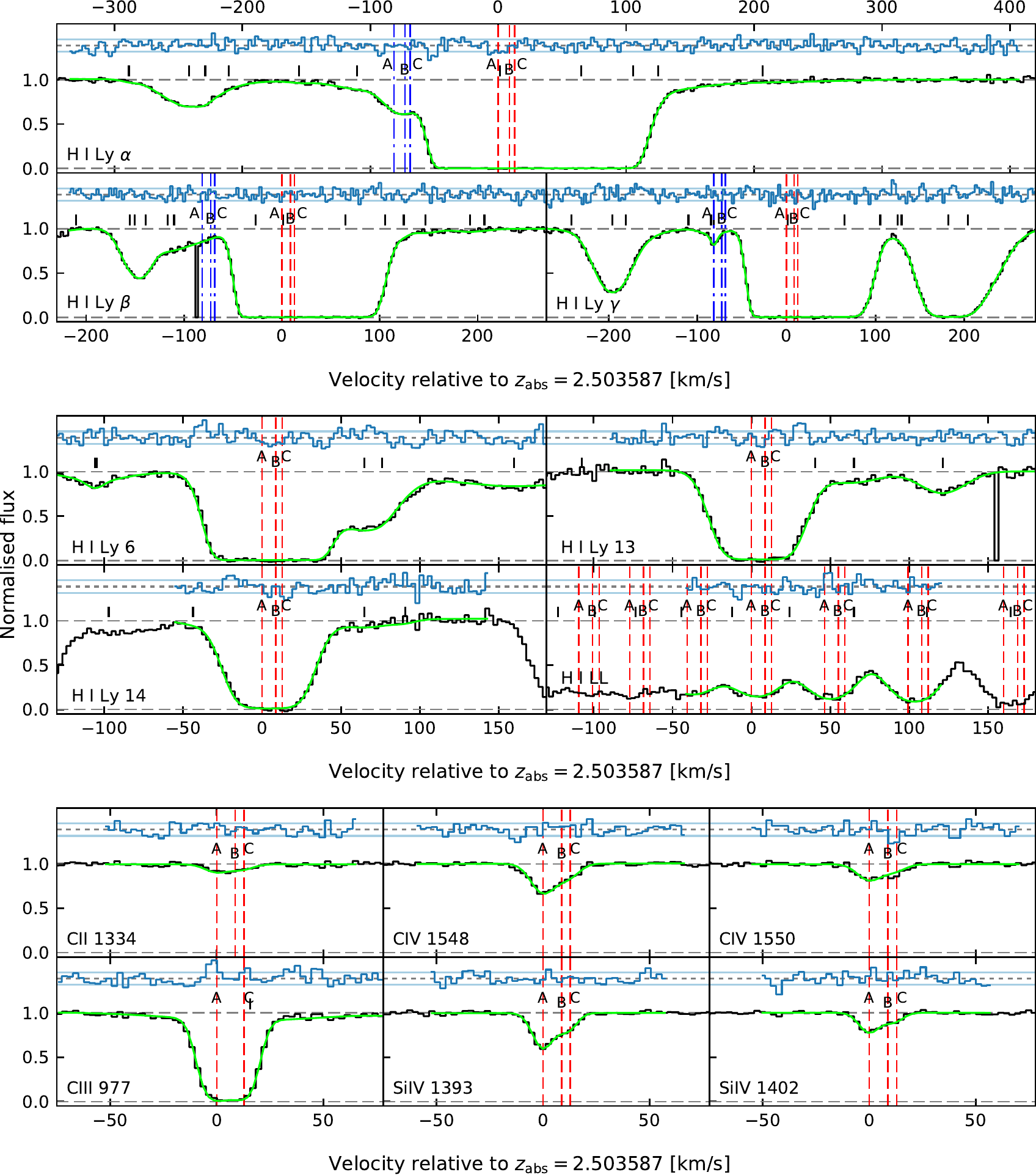}
\caption{Overplotted with the C1$\times$2 spectrum.}
\end{subfigure}
\caption{The overall absorption profile for model 6a (green solid line), the spectrum (black histogram) and residuals with $1\sigma$ confidence range (blue histogram above each fit).  Figure (a) illustrates the C1$\times2$ data and Figure (b) shows the C5$\times1$ data. The two co-added spectra illustrated are the highest S/N. However all four co-added spectra were fitted. The transition corresponding to each panel is indicated in the bottom left corner. The main components A, B, and C are indicated by vertical red dashed lines with letters above. The upper portion of the Figure illustrates three H\,{\sc i} panels, in which the positions of the corresponding deuterium transitions are indicated by blue vertical dash-dotted lines. All the Lyman limit transitions are shown in a single panel (H\,{\sc i}~LL) centred at Ly\,23. The ticks above each fit indicate absorption by additional H\,{\sc i} clouds included in the model. The x-axis is the velocity offset from component A. The absence of component B in C\,{\sc iii} 977 is because {\sc vpfit} iteratively reduced its column density until it fell below the acceptance criterion. The sharp drops of the flux at H\,{\sc i} Ly\,$\beta$ and Ly\,13 of the C1$\times2$ spectrum and C\,{\sc iv} 1550 of the C5$\times1$ spectrum are pixels clipped due to cosmic rays.}
\label{fig:2_504_model_6}
\end{figure*}
\begin{figure*}
\ContinuedFloat 
\begin{subfigure}{\linewidth}
\centering
\includegraphics[width=\linewidth]{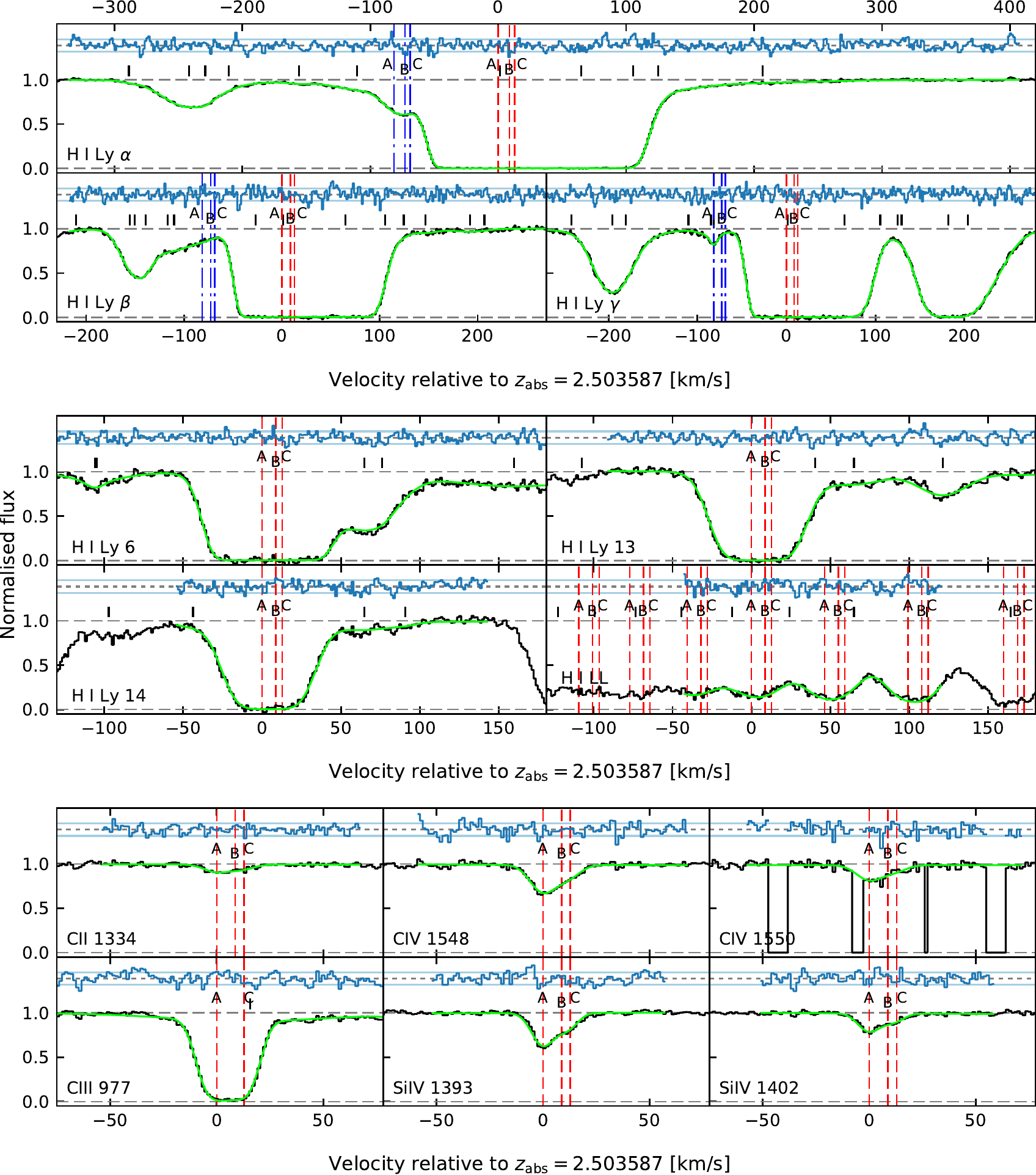}
\caption{Overplotted with the C5$\times$1 spectrum.}
\end{subfigure}
\end{figure*}

\subsection{Model comparison. Preferred number of velocity components}

To establish the number of absorbing components favoured by the data, we use an Akaike information criterion \citep{Akaike1974} corrected for finite sample sizes \citep[e.g.][]{Burnham2002}:
$$
\text{AICc} = \chi^2 + 2p + \frac{2p(p+1)}{n-p-1}
$$
where $n$ and $p$ are the sample size and the number of model parameters, respectively. Unlike to $\chi^2$, AICc includes a penalty for including too many parameters (i.~e. over-fitting). The model with the smallest AICc is the one preferred by the observational data. One can of course only compare AICc values when using the same datasets, i.\,e. we can use AICc to compare models 1 through 3 or models 4 and 5 but not to compare e.\,g. models 3 and 5 (since an additional metal transition is used in the latter). We use the following scale of levels of empirical support of a given model \citep{Burnham2002}: once a model with the smallest AICc (AICc$_{\rm min}$) is found, models that have $\Delta\text{AICc} = \text{AICc} - \text{AICc}_{\rm min}$ greater than 10 are considered as having no support by the data and may be discarded. Models with $4<\Delta\text{AICc}<7$ and $0<\Delta\text{AICc}<2$ have ``considerably less'' and ``substantial'' empirical support, respectively.

Table~\ref{tab:d2h_results} contains AICc values for models 1a to 6a
and Table~\ref{tab:d2h_results_broad} gives the results for 1b to 6b.  AICc differences are given between the given and the smallest ones within each of three model subsets: models 1-3, models 4-5, and model 6.  We have avoided inter-comparing models a and b (in terms of AICc) because models b clearly have larger $\chi^2$ values.  However, this does not rule out models of this sort, i.e. models including a broad component to fit the Ly\,$\alpha$ wings since these models could be refined further, i.e. by including further blends.

Referring to Table \ref{tab:d2h_results} first, all six models appear to give an acceptable $\chi^2$. However, based on the AICc values, a one-component model (model 1a) is strongly disfavoured ($\Delta\text{AICc}=18.1$) with respect to the lowest AICc model (model 3a). Apart from this statistical test, multiple structures are seen in all the metals making a one-component model (model 1a) very unlikely. Two-component models 2a and 4a have ``considerably less'' empirical support with respect to the three-component models 3a ($\Delta\text{AICc}=4.6$) and 5a ($\Delta\text{AICc}=7.2$), respectively. With this simple statistical test, we conclude that three-component models are favoured by our data. 

Given the three three-component models (models 3a, 5a and 6a) explored, what are the arguments, if any, to favour one above another?  As Table \ref{tab:d2h_results} and Figure \ref{fig:2.504_blend_profile} indicate, the interloping H\,{\sc i} line blends particularly badly with D\,{\sc i} for model 3a, leading to the large uncertainty on D/H. Hence, if we were to opt for a model that made no use of any metal line information, we would then have to conclude that without metals, one cannot extract D/H with high precision from this absorption system. However, metals are detected and it therefore is sensible to make optimal use of them to derive the best possible information on velocity structure. 

Of the three-component models, model 6a uses the most metal lines to estimate the velocity structure and constrain gas temperature and turbulent motions.  It is appropriate to use a mix of low and higher ionisation metal species to constrain the velocity structure because H\,{\sc i} column densities are fitted as unconstrained parameters. If any velocity component has ionisation conditions such that all hydrogen is ionised, the H\,{\sc i} column density in this component would be iteratively reduced below the threshold and the corresponding H\,{\sc i} component would subsequently be rejected by {\sc vpfit}.  In other words, we do not require that any particular velocity component necessarily exhibits H\,{\sc i} absorption.  If other velocity components were required that do not have corresponding metal lines, this would be indicated in the appropriate places by poor normalised residuals.  This is not seen in the analysis.

\begin{figure}
\center
\includegraphics[width=\linewidth]{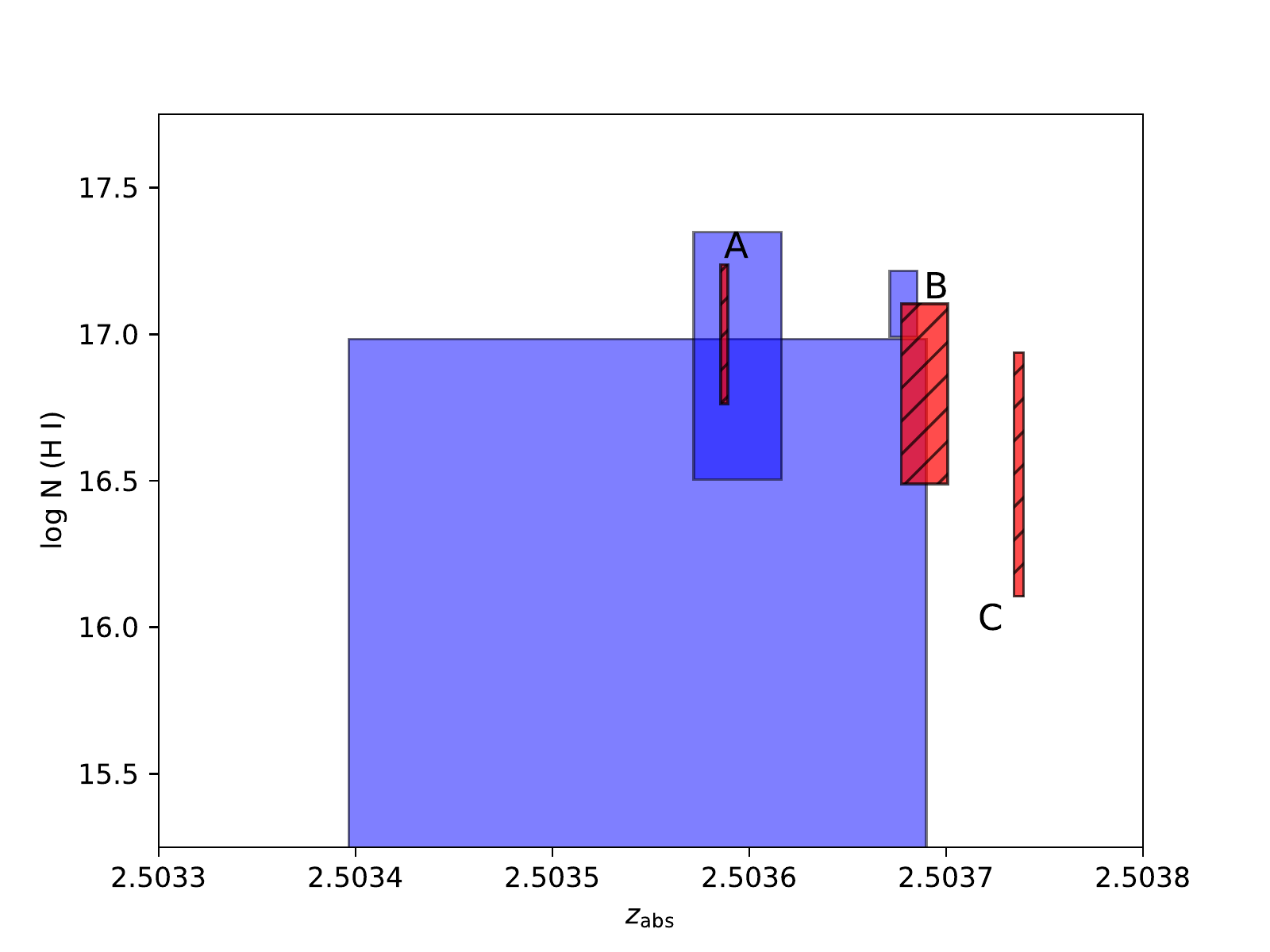}
\caption{Comparison between the $N$\,(H\,{\sc i}) and $z$ parameter estimates for models 3a (derived by fitting to the Lyman series alone) and 6a (with metals used). Each rectangle illustrates the parameter range for one absorption component (three main components in each model). The red hatched rectangles illustrate model 6a.  The labelling A, B and C assumes that we can associate the largest blue (continuous shaded) rectangle with the right red (hatched) rectangle, and that the remaining 2 pairs of rectangles are matched to their closest neighbour.  The two models are shown to be consistent with each other, although model 3a clearly is considerably less-well determined.}
\label{fig:model3vs6}
\end{figure}

We can compare models 3a and 6a. Each has the same number of absorbing components, but no metals were used in fitting model 3a (apart from having used the metal lines to show that all three main components are thermally broadened, see Section~\ref{sec:kin_structure}). Fig.~\ref{fig:model3vs6} illustrates the parameter estimates and associated uncertainties for models 3a and 6a, showing the locations of the 3 main components, A, B and C, for both models in the log\,$N$(H\,{\sc i}$) - z$ plane. Models 3a and 6a are consistent in both $N$ and $z$, provided we associate the largest blue (uniformly-shaded) rectangle with component C in model 6a. From Fig.~\ref{fig:d2h_results}, although the D/H error bars for models 3a and 6a overlap, the model 3a error bar is huge compared to model 6a, with a 0.3 dex difference for D/H (see Table \ref{tab:d2h_results}).  The much larger uncertainty on D/H in model 3a is because the total D\,{\sc i} column density is poorly constrained, and this is because of the dramatically worse constraint on kinematic structure available when fitting the Lyman series alone. Out of models 1a to 6a, we therefore argue that 6a is the most compelling.

We now turn to the second set of models, summarised in Table \ref{tab:d2h_results_broad} and illustrated (for Ly\,$\alpha$) in Fig.~\ref{fig:2.504_blend_broad_profile}.  Models 1b to 6b were motivated by the apparent symmetry of the residual absorption seen in both wings of the Ly\,$\alpha$ profile, hinting at the possible presence of one very broad interloper.  Such a high $b$-parameter would be unusual and inconsistent with the general $b$ distribution.  Nevertheless, the data itself suggests the possibility, as discussed in Section \ref{sec:2.504_broad_blending}.  The one broad interloper in these models replaces four narrow interlopers used in models 1a to 6a.  The interesting aspect of these models is that the broad interloper no longer causes such severe degeneracy between a narrow blend and the deuterium feature.  This is reflected by the dramatically smaller uncertainties on D/H for this set of models compared to models 1a-6a.

Whilst all six models, 1b to 6b, yield consistent D/H values, for the same reasons outlined previously, we identify model 6b as the most reliable in this set.  Since models 6a and 6b make use of the same datasets, we can compare AICc values, and see that, formally, 6a is the preferred model.  However, there is some degree of arbitrariness in that choice as it would be relatively straightforward to make slight refinements to the model (i.e. by including one additional interloper) to reduce both $\chi^2$ and AICc.  For that reason we think it reasonable to identify model 6b as an acceptable solution and use it to estimate a systematic contribution to the final error budget and our final D/H.

\subsection{Systematic uncertainty and the final D/H estimate}

In order to estimate the systematic error associated with the choice of the velocity structure (i.e. the blending model), we adopt a systematic error corresponding to half the difference between the model 6a and 6b results, $\sigma_{\rm sys}(\text{log D/H}) = 0.08$.  Since this gives a crude estimate of the total rather than the systematic error, using this as the systematic component is conservative.  This additional error is random in nature, so we add it in quadrature to the statistical error for models 6a and 6b.  Our final D/H for the LLS analysed in this paper is a weighted mean of both values,
$$ \text{log(D/H}) = -4.606 \pm 0.066$$
or
$$ \text{D/H} = 2.48^{+0.41}_{-0.35} \times 10^{-5}$$
in excellent agreement with the CMB value from \cite{Coc2015} of $\text{D/H}= (2.45\pm0.05)\times 10^{-5}$.

\section{Discussion}

\subsection{Comparison with previous measurement of \citetalias{Burles1998b}} 
\label{sec:comparison_BT}

In the previous analysis of this LLS, \citetalias{Burles1998b} explored six different models. However, no metals were used to constrain the velocity structure. Their preferred model is a three-component model, as is ours (although we note that their model and ours differs not only in terms of additional blends impacting on the D line, as we have described in Section \ref{sec:2.504_multi_blending}, but also elsewhere --- compare \citetalias{Burles1998b}'s fig.~8 with our Fig.~\ref{fig:2_504_model_6}).

\citetalias{Burles1998b} give a value of log\,D/H~$= -4.40^{+0.06}_{-0.08}$.  In some of our models, the fitting process positioned a blend close to the D\,{\sc i} line, degrading the precision with which one can measure $N$(D\,{\sc i}) and hence D/H. \citetalias{Burles1998b}'s models do not include this blend.  This is why our final error D/H estimate is not a dramatic improvement on \citetalias{Burles1998b}'s, despite the far better signal to noise. As discussed in Section \ref{sec:2.504_multi_blending}, if we artificially fix the parameters of this blend, the severe impact on the apparent D/H precision becomes apparent; for our model 3a we would derive an uncertainty on log~D/H of $0.027$ and for model 6a, an uncertainty on log~D/H of $0.014$. 

\subsection{The primordial deuterium abundance from a sample of measurements}
\label{sec:primordial_d2h}

Table \ref{tab:d_sample} presents 15 D/H measurements from the literature.  This list is the same as the 13 listed in table 4 of \cite{RiemerSorensen2017} plus two further systems: \cite{Srianand2010} and \cite{Cooke2016b}.  It is well known that the apparent scatter in D/H measurements exceeds that expected on the basis of the statistical error estimates.  Table \ref{tab:d_sample} includes two ``low'' D/H estimates that one might opt (subjectively) to exclude as being deviant: \cite{Srianand2010} and \cite{Pettini2001}.  However we choose here to apply a modified Least Trimmed Squares procedure \citep[LTS;][]{Rousseeuw1984} to estimate D/H from the sample, which allows us to avoid such subjective decisions as to which measurements are ``good'' and which are unreliable.

The procedure we adopt is as follows.  As an initial step, we adopt an LTS trimming fraction of 0.85, i.e. we allow LTS to discard 15\% of the points (in this case 2). We thus allow LTS to select the ``best'' $k=13$ from $n=15$ values.  LTS identifies entries 7 \citep{Srianand2010} and 15 \citep{Pettini2001} in Table \ref{tab:d_sample} as the most discrepant. To allow for excess scatter in the sample, above the statistical error, at each LTS iteration (that is, for each combination of $k$ from $n$), we iteratively increase all error bars by adding an additional term in quadrature, $\sigma_{\rm rand}$, such that the normalised $\chi^2$ over all points is unity.  The ``successful'' combination of 13 points is taken to be that having the smallest value of $\sigma_{\rm rand}$.  Interestingly, this value turns out to be zero, i.e. the scatter amongst the remaining 13 values is consistent, on average, with the published errors.  The final result, a weighted mean of these 13 D/H values, is
\begin{equation}
(\text{D/H})_{\rm p} = (2.545 \pm 0.025) \times 10^{-5}.
\label{eq:primordial_d2h}
\end{equation}

The value above, based on a relatively small sample of quasar observations, is seen to be in good agreement with the CMB values derived by \cite{Coc2015}, D/H~$= (2.45\pm0.05)\times 10^{-5}$, and by \cite{Marcucci2016}, D/H~$(2.49 \pm 0.03 \pm 0.03)\times10^{-5}$.   The differences between these two CMB values arise from the different nuclear reaction rates used to determine the primordial abundances \citep[see][for a comprehensive discussion]{Cooke2016b,RiemerSorensen2017b}.  The precision of the quasar data is now approaching the point such that observations based on quasar absorption systems may provide a method of determining nuclear rates. 

\begin{table*}
\centering
\caption{D/H measurements used to estimate $\Omega_b h^2$.  A Least Trimmed Squares procedure was used to eliminate outliers (rejection of 2 points, being 15$\%$ of the total set of 15).  LTS rejected the measurements of J1337+3152 and Q2206-199.  See Section \ref{sec:primordial_d2h} for details.}
\label{tab:d_sample}
\begin{tabular}{clcccrllc}
\hline
& Quasar	& $z_{\rm em}$ & $z_{\rm abs}$ & log$N$(H\,{\sc i}) & [X/H], X & D\,{\sc i}/H\,{\sc i} ($\times10^5$) & References&\\
\hline
& HS 0105$+$1619 & 2.65 & 2.537 & $19.426 \pm 0.006$ & $-1.77$ O & $2.58^{\,+\,0.16}_{\,-\,0.15}$ & \cite{Cooke2014} & \parbox[c][4.5mm][c]{0mm}{}\\
& J0407$-$4410 & 3.02 & 2.621 & $20.45\pm 0.10$		& $-1.99$ O  & $2.8^{+0.8}_{-0.6}$			& \cite{Noterdaeme2012} & \parbox[c][4.5mm][c]{0mm}{} \\
& Q0913$+$072 & 2.79 & 2.618 & $20.312 \pm 0.008$ & $-2.40$ O & $2.53^{\,+\,0.11}_{\,-\,0.10}$ & \cite{Cooke2014} & \parbox[c][4.5mm][c]{0mm}{}\\
& Q1009$+$2956 & 2.63 & 2.504 & $17.362\pm 0.005^a$		& $-2.5$ Si$^b$  & $2.48^{+0.41}_{-0.35}$			& This work & \parbox[c][4.5mm][c]{0mm}{} \\
& J1134$+$5742 & 3.52 & 3.411 & $17.95\pm 0.05$		&$<-4.2$ Si  & $2.0^{+0.7}_{-0.5}$			& \cite{Fumagalli2011} & \parbox[c][4.5mm][c]{0mm}{} \\
& Q1243$+$3047 & 2.56 & 2.526 & $19.761 \pm 0.026$		& $-2.77$ O  & $2.39\pm0.08$		& \cite{Cooke2018} & \parbox[c][4.5mm][c]{0mm}{} \\
& J1337$+$3152 & 3.17 & 3.168 & $20.41\pm 0.15$		& $-2.68$ Si &	$1.2^{+0.5}_{-0.3}$			& \cite{Srianand2010} & \parbox[c][4.5mm][c]{0mm}{} \\
& SDSS J135803.97$+$034936.0 & 2.89 & 2.853 & $20.524 \pm 0.006$ & $-2.80$ O & $2.62\pm 0.07$				   & \cite{Cooke2016b}  &  \parbox[c][4.5mm][c]{0mm}{}\\
& J1358$+$6522 & 3.17 & 3.067 & $20.495 \pm 0.008$ & $-2.33$ O & $2.58\pm 0.07$				   & \cite{Cooke2014}	 &  \parbox[c][4.5mm][c]{0mm}{}\\
& J1419$+$0829 & 3.03 & 3.050 & $20.392 \pm 0.003$ & $-1.92$ O & $2.51\pm 0.05$				   & \cite{Cooke2014} & \parbox[c][4.5mm][c]{0mm}{}\\
& J1444$+$2919 & 2.66 & 2.437 & $19.983\pm 0.010$	& $-2.04$ O  & $1.97^{+0.33}_{-0.28}$		& \cite{Balashev2016} & \parbox[c][4.5mm][c]{0mm}{} \\
& J1558$-$0031 & 2.82 & 2.702 & $20.75 \pm 0.03$   & $-1.55$ O & $2.40^{\,+\,0.15}_{\,-\,0.14}$ & \cite{Cooke2014} & \parbox[c][4.5mm][c]{0mm}{}\\
& Q1937$-$1009$^c$ & 3.79 & 3.256 & $18.09\pm 0.03$		& $-1.87$ O  & $2.45^{+0.30}_{-0.27}$		& \cite{RiemerSorensen2015} & \parbox[c][4.5mm][c]{0mm}{} \\
&				&      & 3.572 & $17.925\pm 0.006$ 	& $-2.26$ O  & $2.62\pm 0.05$				& \cite{RiemerSorensen2017} & \parbox[c][4.5mm][c]{0mm}{}\\
& Q2206$-$199 & 2.56 & 2.076 & $20.436\pm 0.008$	& $-2.04^d$ O& $1.65\pm 0.35$				& \cite{Pettini2001} & \parbox[c][4.5mm][c]{0mm}{} \\
\hline
&\multicolumn{5}{l}{CMB prediction} & $2.45\pm0.05$ & \cite{Coc2015} \\
\hline
\end{tabular}

\small{$^a$Mean of the values for models 6a and 6b.}
\small{$^b$Measured by \citetalias{Burles1998b}.}
\small{$^c$There are two absorption systems on the sight-line towards Q1937$-$1009 with identified D\,{\sc i} lines.}
\small{$^d$Measured by \cite{Pettini2008a}.}
\end{table*}

\subsection{The baryon density of the Universe}
\label{sec:omegadh2}

Using the fitting formula for the BBN calculations \citep[][private communication]{Coc2015}:
$$
(\text{D/H})_{\rm p} = (2.45 \pm 0.04) \times 10^{-5}\left(\frac{\Omega_{\rm b}h^2}{0.02225}\right)^{-1.657},
$$
we infer the baryon density of the Universe, the statistical uncertainty, and the uncertainty from nuclear data parameters:
$$
\Omega_{\rm b}h^2 = 0.02174 \pm 0.00013_{\text{\sc qso}} \pm 0.00021_{\rm nucl}.
$$
The overall uncertainty is now dominated by the nuclear data imprecision rather than low statistics of the high redshift quasar data. Comparing our result to the Planck 2015 TT$+$lowP$+$lensing value \citep{Planck2015} (see their table 4),
$$
\Omega_{\rm b}h^2 = 0.02226 \pm 0.00023
$$
i.e. there is a a marginal tension at the 1.6$\sigma$ level with the result we present in this paper.

\section{Conclusions}

The Lyman limit system at $z_{\rm abs}=2.504$ towards Q1009$+$2956 (or J1011$+$2941) was previously used to measure the primordial deuterium abundance by \citetalias{Burles1998b}.  It has long been considered one of the more robust measurements.  We embarked on a re-measurement of this system because of newer, far higher signal to noise, data.  Whilst our {\it a priori} expectation was a correspondingly more precise measurement, the higher signal to noise data in fact revealed a more complex velocity structure than had previously been thought.  The consequence is lower precision than expected.  Further, the higher signal to noise of the newer data (up to 147 per pixel in the continuum level of Ly\,$\alpha$, compared to a peak signal to noise of about 60 in the \citetalias{Burles1998b} data) requires us to consider a broader range of absorption system models.  Taking into account both statistical uncertainties and the additional uncertainties associated with velocity structure ambiguity, our final result for this LLS is
$$ \text{D/H} = 2.48^{+0.41}_{-0.35} \times 10^{-5}$$
in excellent agreement with the CMB value.

A weighted mean of 13 D/H values from the literature (including the result reported here) gives a primordial D/H value of
$$
\left(\text{D/H}\right)_{\rm p} = (2.545 \pm 0.025) \times 10^{-5}.
$$
This leads to a baryon density of the Universe of
$$
\Omega_{\rm b}h^2 = 0.02174 \pm 0.00013_{\text{\sc qso}} \pm 0.00021_{\rm nucl}
$$
marginally inconsistent with the Planck CMB data. Further quasar data as well as experimental improvement in the nuclear data are required to establish whether the marginal discrepancy is a random fluctuation or has a physical origin.

\section*{Acknowledgements}

We thank John O'Meara, Scott Burles, David Tytler and Chris Gelino for their feedback about HIRES data used by \citetalias{Burles1998b}; 
Tristan Dwyer for running his code identifying absorption systems for us; Serj Balashev and Michael Wilczynska for useful discussions; an anonymous referee for valuable comments. 
EOZ is supported by an Australian Government Research Training Program (RTP) Scholarship.
EOZ is grateful for the hospitality of the Institute of Theoretical Astrophysics at the University of Oslo during his visit in January-February 2016.
JKW thanks the Centre for Mathematical Sciences in Cambridge for an enjoyable and productive Sabbatical visit.
This work is based on observations collected with the Keck Observatory Archive (KOA), which is operated by the W.\,M. Keck Observatory and the NASA Exoplanet Science Institute (NExScI), under contract with the National Aeronautics and Space Administration. A significant part of the analysis and figures were done using publicly available python packages: matplotlib \citep{Hunter2007}, numpy \citep{vanderWalt2011}, scipy \citep{Oliphant2007}, astropy \citep{Astropy2013}, barak\footnote{\url{https://pypi.python.org/pypi/Barak/0.3.2}} by Neil Crighton, qscan \citep{qscan2017}. This research has made use of NASA's Astrophysics Data System. EOZ is also grateful to GitHub for the Student Developer Pack.




\bibliographystyle{mnras}
\bibliography{mybibliography} 



\appendix

\section{HIRES data reduction issues}
\label{app:reduction_issues}

\subsection{Wavelength distortions}
\subsubsection{Night-sky line correction}

There is an option in MAKEE to apply an offset to the wavelength array (in velocity space) to correct any possible wavelength shift between the ThAr and science exposures.  
In order to determine the offset which generally seems to be between $0.1-0.4$ pixels (unbinned), MAKEE uses the night sky lines in the background of the object exposure.
The problem arises for the current HIRES detector consisting of three CCDs. Since each of the three CCD exposures are reduced separately, offsets are measured independently. Normally, for wavelength coverage of about $3100-6000$~{\AA} there are enough sky lines to determine the offset on the third (red) CCD chip only. To avoid introduction of an intra-chip velocity offset we used the following approach. We, first, reduced all three CCD separately, giving MAKEE a chance to determine the offset. Then, in case the offset was different for different chips (or if the offset was determined only for the red chip as was usually the case) we calculated a weighted mean value of the offset. Finally, we repeated the reduction specifying the offset manually.

\subsubsection{Air to vacuum correction}
\label{app:air2vac_makee}
As found in \cite{Murphy2001}, MAKEE applies an inaccurate, air to vacuum conversion formula. Instead of using the Edlen formula (as used for the wavelength-calibration ThAr lines), MAKEE adopts a Cauchy dispersion relation \cite[for details see Paragraph 2.9 in][]{Murphy2001}. This produces a wavelength-dependent distortion. To resolve this issue, during the data reduction procedure with MAKEE, we prevented any air to vacuum conversion being applied and instead a correction was applied (at the order-combination stage) in UVES\textunderscore popler, which uses the correct Edlen formula.

\subsection{Blemishes on the CCDs}
\subsubsection{Exposure time dependence of bad regions on the legacy CCD}

It is generally known fact that the legacy ``Tektronix'' CCD of HIRES had a few prominent cosmetic defects such as a large felt-tip pin mark near the center of the CCD \citep{Vogt1994b}, three ``bleeding'' bad regions, a ``hot'' corner and a few bad columns.  MAKEE attempts to mask these regions by default. However, we found that the bleeding regions and the hot corner increase in size in both spatial and dispersion directions with increasing exposure time. Occasionally, they extend outside of the masked regions and affect adjacent parts of the spectrum as well as adjacent orders. To tackle this problem we visually inspected all the legacy exposures and excluded all these affected regions. 

\subsubsection{Bad columns on the new CCD}

In contrast to the legacy CCD of HIRES, the current one (upgraded in August 2004) has less cosmetic defects \citep[see item 7 in][]{Vogt2008}. The most problematic ones are bad partial columns which, in contrast to the legacy CCD, go along the echelle orders. If an echelle order happens to be placed on a bad column, significant part of its spectrum is affected and the absorption line profiles in this region become unreliable. It can also mimic a fluctuation of the continuum. In turn, this can affect the normalization process in the vicinity of this region. In order to avoid any possible bias we identified the problematic regions by visual inspection of the flat-field spectra and then excluded all affected regions.  We managed to identify four bad columns\footnote{There is also a fifth column. However, it is located close to the blue edge of the chip in spatial direction so that no orders are usually extracted from this part of the CCD.} on the first (blue) chip, one bad column on the second (green) chip and two bad columns on the third (red) chip. All these columns start from the blue edge (in dispersion direction) of the chips and cover up to three-quarters of the CCDs.

\subsection{Scattered light}

Most HIRES images of bright objects (quartz lamp, bright star) are affected by scattered light \citep[see, e.g.,][]{Vogt1994b,Vogt2008}. Having visually inspected all the flat-field frames we used in this work using ds9\footnote{\url{http://ds9.si.edu/site/Home.html}}, we identified at least three scattered light features in the legacy data and at least seven features in the current CCD exposures. Occasionally, this light affects the spectrum significantly, in our case --- exposures from 5 November 2004 and 31 March 2005. The scattered light on the first (blue) chip, where the deuterium lines fall, causes an increase of the flux in the flat-field images by up to $10~\%$ introducing a systematic bias during the flat-fielding process.  To avoid any possible bias we completely excluded the affected parts of these spectra from the analysis.

\section{Absorption systems towards Q1009+2956}
\label{app:abs_systems}
In Table~\ref{tab:abs_systems} we provide a list of absorption systems and corresponding species identified in the spectrum of Q1009$+$2956. The technique used in the identification process is described in Section~\ref{sec:poss_cont}.

\begin{table}
\centering
\begin{minipage}{\linewidth}
\caption{List of absorption systems and corresponding species tentatively identified in the spectrum of Q1009$+$2956.}
\label{tab:abs_systems}
\begin{tabular}{@{}cr@{}}
	\hline
	Redshift		& Species$^a$  \\
	\hline
	$-0.0001^b$	& Na\,{\sc i}, Ca\,{\sc ii}, Ti\,{\sc ii} \\
	0.8599		& Fe\,{\sc ii}, Mg\,{\sc ii} \\
	1.1117 		& C\,{\sc iv} (?), Al\,{\sc ii}, Al\,{\sc iii}, Fe\,{\sc ii}, Mg\,{\sc i} (?), Mg\,{\sc ii} \\
	1.1146 		& C\,{\sc iv}, Mg\,{\sc ii} \\
	1.1707 		& C\,{\sc iv} \\
	1.1729 		& C\,{\sc iv}, Al\,{\sc ii} (?), Al\,{\sc iii}, Mg\,{\sc ii} \\
	1.2709 		& C\,{\sc iv}, Al\,{\sc ii}, Al\,{\sc iii}, Mg\,{\sc ii}, Fe\,{\sc ii} \\
	1.5211 		& C\,{\sc iv}, Si\,{\sc iv} \\
	1.5583 		& C\,{\sc iv} (?), Si\,{\sc iv} (?) \\
	1.5986 		& C\,{\sc ii}, C\,{\sc iv}, Si\,{\sc ii}, Si\,{\sc iv}, Al\,{\sc ii}, Al\,{\sc iii}, Fe\,{\sc ii}, N\,{\sc v} (?) \\
	1.9061 		& C\,{\sc iv} \\
	2.1437 		& C\,{\sc iv}, Si\,{\sc iii} (?), Si\,{\sc iv} (?) \\
	2.2062 		& C\,{\sc iv} \\
	2.2073 		& C\,{\sc iv} \\
	2.2533 		& C\,{\sc iii} (?), C\,{\sc iv}, Si\,{\sc ii}, Si\,{\sc iii}, Si\,{\sc iv}, Al\,{\sc ii} (?) \\
	2.3612 		& C\,{\sc iv} \\
	2.4070 		& C\,{\sc ii}, C\,{\sc iii}, C\,{\sc iv}, Si\,{\sc ii}, Si\,{\sc iii}, Si\,{\sc iv}, Al\,{\sc ii}, Al\,{\sc iii}, Fe\,{\sc ii}, Ni\,{\sc iii} (?) \\
	2.4271 		& C\,{\sc iii}, C\,{\sc iv}, Si\,{\sc iv} \\
	2.4290 		& C\,{\sc ii}, C\,{\sc iii}, C\,{\sc iv}, Si\,{\sc ii}, Si\,{\sc iii}, Si\,{\sc iv}, Al\,{\sc ii}, Al\,{\sc iii}, Ni\,{\sc iii} (?) \\
	2.5036 		& C\,{\sc ii}, C\,{\sc iii}, C\,{\sc iv}, Si\,{\sc ii}, Si\,{\sc iii}, Si\,{\sc iv} \\
	2.5533 		& C\,{\sc iv} \\
	2.5726 		& C\,{\sc iii} (?), C\,{\sc iv} \\
	2.6059 		& C\,{\sc iv}, Al\,{\sc ii} (?) \\
	2.6490 		& C\,{\sc iii}, C\,{\sc iv}, Si\,{\sc iv}, N\,{\sc v} \\
	\hline
\end{tabular}

\small{$^a$Question marks indicate possible detections. Usually, it corresponds to species with only one transition available which is also blended significantly.}\\
\small{$^b$Absorption by the interstellar medium.}
\end{minipage}
\end{table}

\section{Description of the supplementary files}
\label{app:suppl_files}

All the {\sc vpfit} files with the best-fit parameters and their uncertainties for each tested model, the co-added spectra and other supplementary files are available in online supplementary materials and on GitHub (the links are given in Section~\ref{sec:2_504_fit_results}).

All the supplementary files come in four separate folders: \texttt{data/}, \texttt{vpfit/}, \texttt{abs\_syst/} and \texttt{voffset/}:
\begin{itemize}
\item \texttt{data/} contains four text files representing four co-added spectra (see Section~\ref{sec:offset}). Each text file has three columns: wavelength in vacuum-heliocentric frame, normalised flux and $1\sigma$ error. Clipped pixels have a negative value assigned to their errors.
\item \texttt{vpfit/} contains an \texttt{atom.dat} file with atomic data used, \texttt{vp\_setup.dat} \textsc{vpfit} setting files and resulted \textsc{vpfit} output files for all the considered models. The latter contain all parameters for each model. Numeration of the models matches the one given in Sections~\ref{sec:kin_structure}, \ref{sec:2.504_multi_blending} and \ref{sec:2.504_broad_blending}. In addition, the \texttt{metals/} sub-folder contains \textsc{vpfit} output and \texttt{vp\_setup.dat} files for the thermal and turbulent fits of metal lines mentioned in Section~\ref{sec:kin_structure}. We refer an interested reader to the \textsc{vpfit} manual\footnote{Available at \url{https://www.ast.cam.ac.uk/~rfc/vpfit.html}.} for details on the format of these files. 
\item \texttt{voffset/} contains a file with velocity offsets in km s$^{-1}$ for each individual exposure with respect to the one obtained on 29 March 2008 (starting time 06:40, UT).  See Section~\ref{sec:offset} for description of how these offsets were calculated. Names of the exposures are given in the KOA format\footnote{Described at \url{https://www2.keck.hawaii.edu/koa/public/faq/koa_faq.php\#U9}.}.
\item \texttt{abs\_syst/} contains a list of identified species per absorption system described in Section~\ref{sec:poss_cont}. This is just an electronic version of Table~\ref{tab:abs_systems}. 
\end{itemize}


\bsp	
\label{lastpage}
\end{document}